\documentclass[english]{ThesisMASTAT}
\pdfoutput=1

\newcommand{\code}[1]{{\normalfont\ttfamily #1}}

\newcommand{\E}{\mathrm{E}}
\newcommand{\Var}{\mathrm{Var}}
\newcommand{\Cov}{\mathrm{Cov}}

\newcommand{\overbar}[1]{\mkern 1.5mu\overline{\mkern-1.5mu#1\mkern-1.5mu}\mkern 1.5mu}

\title{Fitting Probabilistic Index Models on Large Datasets}
\author{Han Bossier}

\promotor{Prof. Dr. Olivier Thas}
\copromotor{Prof. Dr. Jan De Neve}
\tutor{Dr. Gustavo Amorim}

\faculty{Sciences}

\academicyear{2016 - 2017}

\begin{document}
	\frontmatter
	 \pagestyle{empty} 
	\chapter*{Foreword} 

I would like to thank Gustavo Amorim, Jan De Neve and Olivier Thas for providing me the opportunity to work on the interesting topic of Probabilistic Index Models. Gustavo has helped me understanding PIMs, gave the idea to try out non uniform subsampling and gave references to the bootstrap literature. Jan has provided me the calculation for the adjusted sandwich variance estimator, revised parts of the thesis and gave explanation and critical comments on the real data analysis. Olivier has provided the initial idea for subsampling and revised parts of this thesis. \\

In this dissertation, we use anonymized real data of Przybylski and Weinstein (2017), available through the Open Science Framework \url{(https://osf.io/82ybd/}). Enjoy!


	\glsaddall
	
\clearpage{\thispagestyle{empty}\cleardoublepage}

\renewcommand{\contentsname}{Table of Contents} 
\tableofcontents

\clearpage{\thispagestyle{empty}\cleardoublepage}

%
	
	\chapter*{Abstract} 

Recently, \cite{Thas2012} introduced a new statistical model for the probability index. This index is defined as $P(Y \preceq Y^{*}|\mathbf{X}, \mathbf{X^{*}})$ where $Y$ and $Y^*$ are independent random response variables associated with covariates $\mathbf{X}$ and $\mathbf{X^{*}}$. The probabilistic index model consists of a link function which defines the relationship between the probabilistic index and a linear predictor. Crucially to estimate the parameters of the model, a set of \textit{pseudo-observations} is constructed. For a sample size $n$, a total of $\frac{n(n-1)}{2}$ pairwise comparisons between observations is considered. Consequently for large sample sizes, it becomes computationally infeasible or even impossible to fit the model as the set of \textit{pseudo-observations} increases nearly quadratically. In this dissertation, we provide two solutions to fit a probabilistic index model. \\
The first algorithm consists of splitting the entire data set into unique \textit{partitions}. On each of these, we fit the model and then aggregate the estimates. A second algorithm is a subsampling scheme in which we select $K \ll n$ observations without replacement and after $B$ iterations aggregate the estimates. In Monte Carlo simulations, we show how the partitioning algorithm outperforms the latter algorithm. The results of the partitioning algorithm show a nearly unbiased estimator for the parameters in the model as well as for the variance estimator. The empirical coverages of the 95\% confidence intervals are close to their nominal level. Furthermore, the sample distribution of the estimates approximates the normal distribution. \\
We illustrate the partitioning algorithm and the interpretation of the probabilistic index model on a real data set \citep{Przybylski2017} of $n = 116,630$ where we compare it against the ordinary least squares method. By modelling the probabilistic index, we give an intuitive and meaningful quantification of the effect of the time adolescents spend using digital devices such as smartphones on self-reported mental well-being. We show how moderate usage is associated with an increased probability of reporting a higher mental well-being compared to random adolescents who do not use a smartphone. On the other hand, adolescents who excessively use their smartphone are associated with a higher probability of reporting a lower mental well-being than randomly chosen peers who do not use a smartphone. We are able able to fit this model in a reasonable time. \\
Finally we discuss the results of the simulation study and the application. 


\newacronym{b}{\textit{B}}{amount of resampling iterations}
\newacronym{ci}{CI}{Confidence Interval}
\newacronym{iid}{i.i.d.}{identically and independent distributed}
\newacronym{k}{\textit{K}}{number of subsampled observations}
\newacronym{ols}{OLS}{Ordinary Least Squares}
\newacronym{pi}{PI}{Probabilistic Index}
\newacronym{pim}{PIM}{Probabilistic Index Model}
\newacronym{s}{\textit{S}}{amount of data partitions}
 
\printglossary[type=\acronymtype, title={List of Abbreviations}] 

	\mainmatter 
	\pagestyle{fancy} 
	\chapter{Introduction}

\section{Setting}
Very often in science, one is interested in describing and quantifying a relation between a predictor $X$ and a response or outcome $Y$. For example in chapter \ref{Chapter:real_data}, we will look at the association between the amount of time spent using digital devices and self-reported mental well-being in adolescents. Generally, a data-analyst will describe this relationship in terms of the average response. We will for instance see how an increase of 1 hour time spent using a smartphone during the week is associated with an average decrease of 0.68 points in self-reported mental well being for 15-years old adolescents. However, what is the exact interpretation of 0.68 points mental well-being? We don't know. This is because we measure mental well-being using a questionnaire and hence obtain a response from ordinal scale. More specific, ordinal scales/data represent a rank or order, but do not allow for a relative nor absolute degree of differences. They do not posses a natural interpretation about distances between levels. Alternatively, one could ask what the probability is of subjects who spend one hour with their smartphone reporting a \textit{lower} mental well-being compared to subjects who are not using a smartphone. This is a statement about order. It turns out that this probability is equal to 47.93\%. This quantification of this effect is termed a \gls{pi}. \\
In this dissertation, we focus on a class of models called the \gls{pim} \citep{Thas2012}. These are developed to describe a relation between $X$ and $Y$ in terms of the \gls{pi} \citep[see also:][]{Enis1971, Browne2010, Zhou2008, Tian2008}. However, fitting these models can become problematic (computational wise) if the data set becomes too large. Although this is true for every statistical model, we will see how this is particularly so for \gls{pim}s as the effective sample size for which the model is defined in most cases is equal to $\frac{n(n-1)}{2}$, where $n$ is the original sample size. The goal of this dissertation is twofold. First we will explore easy to implement solutions to fit a \gls{pim} on a data set that is too large. This is defined as the sample size for which, under standard estimation procedure, we are unable to obtain estimates under reasonable computational time (or even at all). Then we will apply the best working solution to a real data set and provide an alternative interpretation to the average response. \\
Before this, we will discuss the \gls{pi} and introduce the \gls{pim}.

\section{Introduction to probabilistic index models}
In this section we give a concise introduction to \gls{pim}s. Later, we will provide more details and discuss the standard estimation procedure. Note that the following overview is primarily based on \cite{Thas2012} and \cite{DeNeve2013}. Some details regarding the underlying theory are omitted as only relevant parts for this dissertation are discussed. \\

To start with, denote a single response variable $Y$ and a $d$-dimensional covariate as $\mathbf{X}$. Let $(Y, \mathbf{X})$ and $(Y^*, \mathbf{X}^*)$ be \gls{iid} with joint density $f_{Y\mathbf{X}}$. The \gls{pi} is defined as
\begin{align}  \label{eqPI}
P(Y \preceq Y^{*}|\mathbf{X}, \mathbf{X^{*}}) := P(Y < Y^{*}|\mathbf{X}, \mathbf{X^{*}}) + \frac{1}{2} P(Y = Y^{*}|\mathbf{X}, \mathbf{X^{*}}).
\end{align}
This index represents an effect of the covariates on the response variable. Note when $Y$ is continuous, then $P(Y = Y^{*}|\mathbf{X}, \mathbf{X^{*}}) = 0$ and equation (\ref{eqPI}) simplifies to $P(Y \preceq Y^{*}|\mathbf{X}, \mathbf{X^{*}}) = P(Y < Y^{*}|\mathbf{X}, \mathbf{X^{*}})$. Furthermore, this index behaves intuitively when $Y$ is ordinal as $P(Y \preceq Y^{*}|\mathbf{X} = \mathbf{x}, \mathbf{X^{*}} = \mathbf{x}) = \frac{1}{2}$. \\

The \gls{pim} is now defined as:
\begin{align} \label{eqPIM}
P(Y \preceq Y^{*}|\mathbf{X}, \mathbf{X^{*}}) := m(\mathbf{X},\mathbf{X^*}; \boldsymbol{\beta}).
\end{align}
Here, $m(.)$ is a function which necessarily has a range of $[0, 1]$ and $\boldsymbol{\beta}$ is a $p$-dimensional parameter vector. Note that $P(Y \preceq Y^{*}|\mathbf{X}, \mathbf{X^{*}}) + P(Y^{*} \preceq Y|\mathbf{X}, \mathbf{X^{*}}) = 1$, which implies that $m(\mathbf{X}, \mathbf{X^*}; \boldsymbol{\beta}) = 1 - m(\mathbf{X^*}, \mathbf{X}; \boldsymbol{\beta})$ and $m(\mathbf{X}, \mathbf{X}; \boldsymbol{\beta}) = m(\mathbf{X^*}, \mathbf{X^*}; \boldsymbol{\beta}) = 0.5$. \\

While model (\ref{eqPIM}) does make restrictions on the conditional distribution of Y given $\mathbf{X}\ (\text{i.e. } f_{Y|\mathbf{X}})$, no full distributional assumptions are made on $f_{Y|\mathbf{X}}$. The model thus requires semiparametric theory for inference on $\boldsymbol\beta$. \\

For clarity, we demonstrate the interpretation of a \gls{pim}. We use an example discussed more in detail in Chapter \ref{Chapter:real_data}. There we look at a data set from \cite{Przybylski2017} which contains $n = 116,630$ adolescents from the UK. Participants were asked to keep record of the time they spent using electronic devices (computer, television, smartphone) during one week. They were also asked to fill in the 14 item self-reported Warwick-Edinburgh Mental Well-Being Scale \citep{Tennant2007}. Higher values indicate a higher psychological and social well-being. Scores range from 14 to 70. Denote $\{MWBI, MWBI^*\}$ as the self reported mental well-being and $\{SMART, SMART^*\}$ as the recorded amount of hours spent using a smartphone during the week for subject $Y$ and $Y^*$ respectively. By using a simple algorithm which we introduce later on, we are able to fit a \gls{pim} on the entire data set. More particularly, we fit:
\begin{align} \nonumber
P(MWBI \preceq MWBI^{*}) = \Phi \left[ \beta(SMART^* - SMART) \right]
\end{align}

We find $\hat\beta = -0.083\ (\text{SE} = 0.0015)$. Testing for $H_0$: $\beta = 0$ with $H_a$: $\beta \neq 0$ reveals a test statistic of $z = -55.12\ (p < 0.001)$. Thus at a significance level of 5\% we reject the null hypothesis. Therefore we state that adolescents from the UK who spend more time using a smartphone are more likely to report a lower mental well-being. Say for instance we compare two randomly chosen adolescents: one who did not use a smartphone during the week and one peer who used his/her smartphone for 3 hours. 
We then obtain an estimated probability of $\text{expit}(3 \times -0.083) = 43.81 \%$ of the first adolescent reporting a lower mental well being. Or in other words, we obtain an estimated probability of $1 - 0.4381 = 56.19\%$ of the adolescent who is using a smartphone for 3 hours to report a lower mental well-being, compared to a peer who is not using a smartphone. The \gls{pi} provides an intuitive and meaningful effect size for ordinal data. \\

\begin{figure}[!t]
\begin{center}
\includegraphics[scale=0.75]{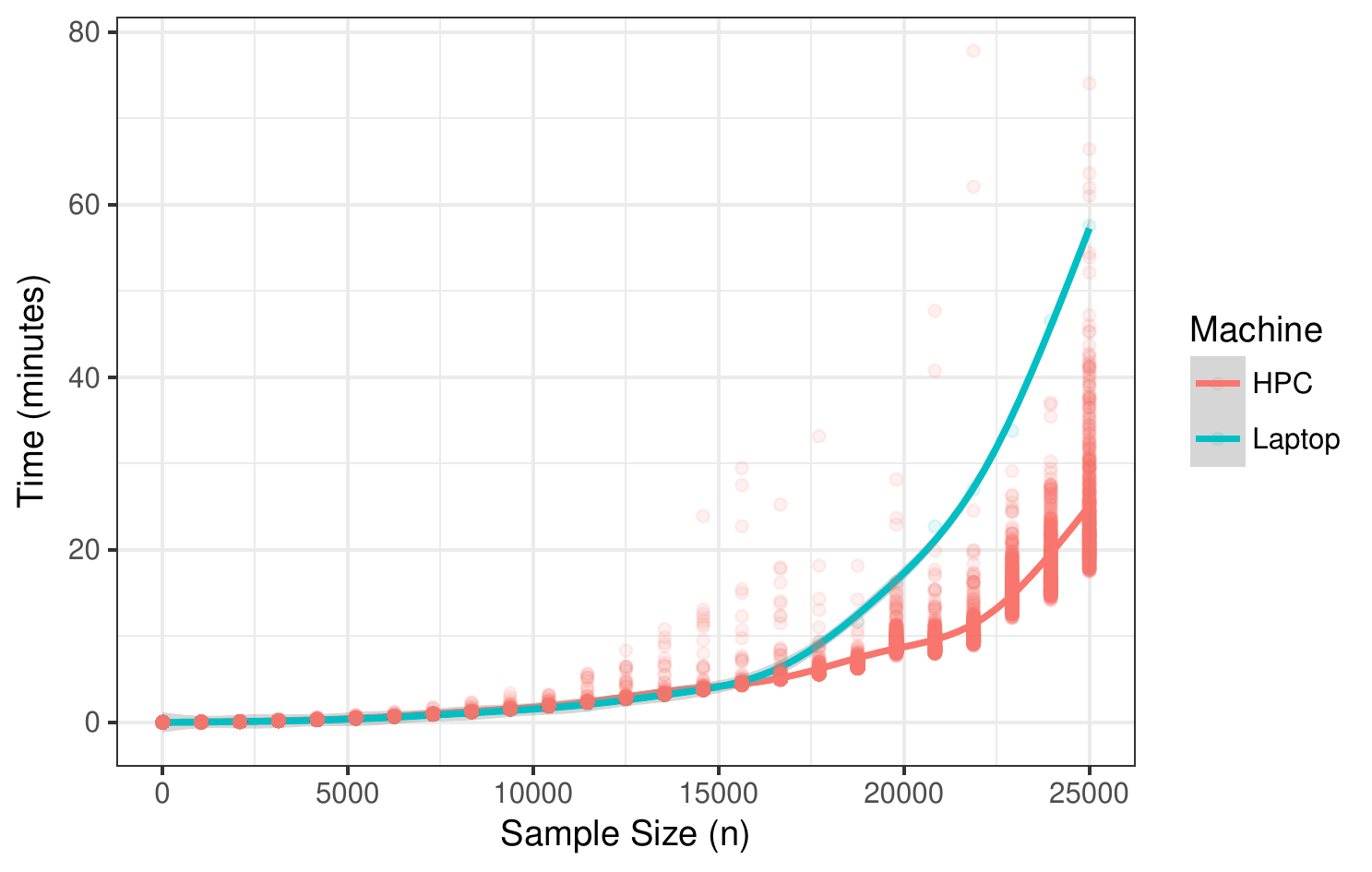}
\caption{Measured time to fit a \gls{pim} on simulated data sets for increasingly larger sample sizes. Each data set contains one predictor. One simulation is run on a single laptop while 500 simulations are calculated with High Performance Computing.}\label{time_to_estimate}
\end{center}
\end{figure}

Before continuing, we stress that a \gls{pim} is not a tool in itself to make causal claims. In the aforementioned data set, we are limited to cross-sectional survey data. Although some covariates are considered in the model later on, there is no guarantee of unmeasured confounders. Nor is there any randomization or temporal order between the predictor and its response. \\
Secondly, the time to fit a \gls{pim} becomes problematic when the sample size increases. To see, define $I(Y \preceq Y^*) := I(Y < Y^*) + 0.5I(Y = Y^*)$, where $I(.)$ is the indicator function evaluating the events $\{Y < Y^*\}$ and $\{Y = Y^*\}$. As shown in \cite{Thas2012}, the expected value of $I(Y \preceq Y^*)$ is given as
\begin{align} \label{exp_PI}
\E{(I(Y \preceq Y^*)| \mathbf{X},\mathbf{X^*})} = P(Y \preceq Y^{*}|\mathbf{X}, \mathbf{X^{*}}).
\end{align}
Equation (\ref{exp_PI}) suggests that the transformed outcome $I(Y \preceq Y^*)$ is calculated over a larger set than $n$ \gls{iid} observations. Instead, model (\ref{eqPIM}) will be fitted on a set of \textit{pseudo-observations} $I_{ij} := I(Y_i \preceq Y_j)$ for all $i,j = 1, \dots n$ for which model (\ref{eqPIM}) is defined. For most \gls{pim}s considered in this dissertation, there are $\frac{n( n- 1)}{2}$ pairwise comparisons which is the set of \textit{pseudo-observations}. This directly suggests how the computational time to fit a \gls{pim} grows nearly quadratic with $n$ as $n$ increases to larger magnitudes. To provide some reference (see Figure \ref{time_to_estimate}), we have simulated data sets with one predictor for increasingly larger sample sizes. On a single laptop, it can take up to 60 minutes to fit a \gls{pim} when $n = 25000$. Using the standard estimation procedure, it would be impossible to fit a \gls{pim} on a data set of $n > 100,000$, which is the focus of Chapter \ref{Chapter:real_data}. \\
The goal of this dissertation is to provide simple solutions to reduce computational time. In the next section, we will introduce two approaches to do so. 

\section{Reducing computational time}
When $n > 2500$, it becomes a necessity to adapt the standard fitting procedure. Our main technique is to reduce the \textit{working} data set. We prefer not to use any transformation nor dimension reduction of the original data, such as done in \cite{Dhillon2013}. The rationale is that, if suitable, an easy to understand solution will be used more often than its complex counterparts. \\

The problem we are facing is certainly not new to the field of (computational) statistics or data science. Some (international) collaborations create large databases of research data. One example is the UK Biobank \citep{Sudlow2015}, which aims to create a database of $n = 500,000$ including neuroimaging data and health measurements. Or vast amounts of data are collected from various automatic sources in a big data era. These create specific challenges to any statistical model. However by using a \gls{pim}, we encounter them in earlier stages. For this reason, it is important to mention that we are not developing a solution to rapidly fit a \gls{pim} in a big data context as there are more challenges related to big data than merely fitting a model \citep{Schutt2013}. These challenges are related to storage, velocity, distributing computational resources, etc. Also, we assume data sets where the amount of predictors is less than $n$. \\
On the other hand, we find inspiration in these research areas to fit \gls{pim}s on larger data sets. For instance, \cite{Wang2016} give an overview on fitting statistical models in a big data context. They categorize three main approaches. These are (in their terminology): \textit{divide and conquer}, \textit{subsampling-based methods} and \textit{online updating}. The last solution is specific for settings in which continuously new observations are added to a database. As this is not relevant here, we will consider the first two types only. \\

Our first algorithm will be called the \textit{single data partitioning} and corresponds to the first approach of \cite{Wang2016}. Briefly, it consists of splitting the entire dataset into $S$ \textit{partitions} which are non-overlapping. Then each partition is analysed separately (preferably in parallel). This is a simple solution, though has a drawback that its computational time is proportional to the sample size. \\
Our second algorithm is called \textit{uniform subsampling}. This algorithm consists of repeatedly sampling $K \ll n$ data points without replacement and then combine the estimates. This approach has been studied in detail in \cite{Politis1994}. The main drawback is that results potentially depend on the choice of $K$ in relation to $n$. \\
These algorithms have a close connection with the method of bootstrap \citep{Efron1994}. Importantly, a standard bootstrap procedure consists of sampling $n$ observations with replacement. The same holds for bootstrap procedures that try to reduce the amount of unique information per bootstrap, such as the bag of little bootstraps \citep{Kleiner2014}. This is not an option for \gls{pim}s as we want to reduce $n$. However, alternatives have been developed such as the \textit{m out of n} bootstrap \citep{Bickel1997, Bickel2008}. Here one samples $m \ll n$ observations with replacement. That said, the goal of a bootstrap method is to estimate the sampling distribution of the test statistic by generating new samples from the underlying population. As we are merely interested in selecting random subsets from the data sample, we choose to implement a subsampling algorithm. \\

In the following chapter, we will introduce the standard estimation procedure for \gls{pim}s and further discuss the two algorithms introduced above. Furthermore, we will provide two approaches to calculate the variance of the estimators obtained through the sampling algorithms. Next we will describe the set-up for our simulation study to investigate the performance of the algorithms. In Chapter \ref{Chapter:SimRes}, we look at the results of this simulation study. Then in Chapter \ref{Chapter:real_data}, we will apply the \textit{single data partitioning} algorithm to a real data set. Finally, results of this dissertation are discussed in Chapter \ref{Chapter:Disc}

	\chapter{Methods}

The code for the simulations and analyses can be found at: \\
\url{https://github.com/HBossier/BigDataPIM}

\section{Probabilistic Index Models}
In this section, we describe the standard estimation procedure to obtain estimates for a \gls{pim}. Before we continue, let us restate the definition of a \gls{pi} for a single response variable $Y$ and a $d$-dimensional covariate $\mathbf{X}$:
\begin{align} \label{Chap2:eqPI}
P(Y \preceq Y^{*}|\mathbf{X}, \mathbf{X^{*}}) &:= P(Y < Y^{*}|\mathbf{X}, \mathbf{X^{*}}) + \frac{1}{2} P(Y = Y^{*}|\mathbf{X}, \mathbf{X^{*}}).
\end{align}
A \gls{pim} is defined as
\begin{align}  \label{Chap2:eqPIM}
P(Y \preceq Y^{*}|\mathbf{X}, \mathbf{X^{*}}) &:= m(\mathbf{X},\mathbf{X^*}; \boldsymbol{\beta}).
\end{align}

Necessarily, $m(.)$ is a smooth function. In order to impose this restriction, \cite{Thas2012} suggest $m(.)$ to be related to a linear predictor $\mathbf{Z}^T\boldsymbol{\beta}$ with $\mathbf{Z} = \mathbf{X^*} - \mathbf{X}$. They propose $m(.)$ to take the form of
\begin{align}
m(\mathbf{X},\mathbf{X^*}; \boldsymbol{\beta}) = g^{-1}(\mathbf{Z}^T\boldsymbol{\beta}),
\end{align}
with $g(.)$ a link function that maps $[0, 1]$ onto the range of $\mathbf{Z}^T\boldsymbol{\beta}$. In this project, we only consider the logit link function, $\log{\left(\frac{p}{1-p}\right)}$ and the probit link function, $g(.) = \Phi^{-1}(.)$ in which $\Phi(.)$ is the cumulative distribution function of the standard normal distribution. \\

Now recall that the expected value of $I(Y \preceq Y^*)$ is given as
\begin{align} \label{Chap2:expPI}
\E{(I(Y \preceq Y^*)| \mathbf{X},\mathbf{X^*})} = P(Y \preceq Y^{*}|\mathbf{X}, \mathbf{X^{*}}).
\end{align}
Upon using equation (\ref{Chap2:expPI}) and (\ref{Chap2:eqPIM}), we can write
\begin{align} \label{exp_link}
\E{(I(Y \preceq Y^*)| \mathbf{X},\mathbf{X^*})} = g^{-1}(\mathbf{Z}^T\boldsymbol{\beta}).
\end{align}
To estimate the parameters of (\ref{exp_link}), one uses the set of \textit{pseudo-observations} $I_{ij} := I(Y_i \preceq Y_j)$ for all $i,j = 1, \dots n$ for which model (\ref{Chap2:eqPIM}) is defined. Note, despite that all $n$ observations are \gls{iid}, the transformed outcome $I(Y \preceq Y^*)$ is no longer independent. Indeed, the \textit{pseudo-observations} posses a \textit{cross-correlation}. If two \textit{pseudo-observations} share the same outcome value, they are in general not mutually independent \citep{DeNeve2013}. \\
To continue, \cite{Thas2012} propose to solve the following estimating equations 
\begin{align} \label{EsEq_1}
U_n(\boldsymbol{\beta}) = \sum_{(i,j) \in I_n} \mathbf{A}(\mathbf{Z}_{ij};\boldsymbol{\beta})[I_{ij} - g^{-1}(\mathbf{Z}_{ij}\boldsymbol{\beta})] = 0,
\end{align}
with $\mathbf{A}(\mathbf{Z}_{ij};\boldsymbol{\beta})$ a $p$-dimensional vector function of the regressors $\mathbf{Z}_{ij}$, which is provided here for reference
\begin{align} \label{EsEq_2}
\mathbf{A}(\mathbf{Z}_{ij};\boldsymbol{\beta}) = \frac{\partial g^{-1}(\mathbf{Z}_{ij}\boldsymbol{\beta})}{\partial \boldsymbol{\beta}} V^{-1}(\mathbf{Z}_{ij};\boldsymbol{\beta}).
\end{align}

The attentive reader will recognize the form of equation (\ref{EsEq_1}) and (\ref{EsEq_2}) as it resembles the class of generalized estimating equations \citep{Zeger1986, Liang1986}. Briefly, two elements are specified. This is (1), the conditional mean of the set of pseudo-observations. And (2), the relationship between the mean and the variance structure \citep{Boos2013, Zeger1986}. The choice of the dependence structure in $V^{-1}(\mathbf{Z}_{ij};\boldsymbol{\beta})$ is a \textit{sparse correlation} structure due to the cross-correlation. This structure has been studied in crossed experimental designs where observations are correlated due to overlapping design factors \citep{Lumley2011}. \\

Now denote the roots of equations (\ref{EsEq_1}) as $\hat{\boldsymbol{\beta}}$. We will further refer to these as the \gls{pim} estimators on the full data set. \cite{Thas2012} demonstrate that as $n \rightarrow +\infty$, $\sqrt{n}(\hat{\boldsymbol{\beta}} - \boldsymbol{\beta})$ converges in distribution to a multivariate Gaussian distribution with zero mean and a positive definite variance-covariance matrix $\boldsymbol{\Sigma}$. The latter can be consistently estimated by the sandwich estimator $\boldsymbol{\hat{\Sigma}}_{\boldsymbol{\beta}}$ (not shown here). \\

Finally consider the following null hypothesis for $\beta_{1}$, the parameter of interest
\begin{align} \label{null}
H_{0}: \beta = \beta_{0},
\end{align}
where $\beta_{0}$ is the corresponds to the null value. Based on the asymptotic normality and the consistent sandwich estimator, Wald type tests can be constructed in the usual way. \\

\section{Sampling algorithms}
As introduced earlier, we consider two algorithms to reduce the computational time when fitting a \gls{pim}. We hypothesize that we will \textit{gain} computational time by working with smaller sets, as the time to fit the model increases nearly quadratically with $n$. \\

The first algorithm is the \textit{single data partitioning} and consists of splitting the complete dataset into smaller non-overlapping subsets, called \textit{partitions}. The second algorithm is \textit{uniform subsampling} and is based on iterating a subsampling scheme in which we draw much less than $n$ observations without replacement within each iteration. \\ 
Both algorithms operate on the original response vector $Y$ (with its covariates), not the set of pseudo-observations. This is chosen as to have independent observations within a sampling iteration or partition, as the pseudo-observations posses a cross-correlation structure. \\
Furthermore, we will denote $\widehat{\boldsymbol\beta}^*$ as the estimators obtained by fitting a \gls{pim} on a subset of the data, while $\widehat{\boldsymbol\beta}$ remains the \gls{pim} estimators corresponding to the full data set. \\
We shall now discuss both algorithms for a \gls{pim} with a one-dimensional predictor $X$ along with proposed calculations for $\widehat{\beta}^*$ (same calculations hold for \gls{pim}s with more predictors).

\subsection{Single data partitioning}
The first algorithm is based on a single partitioning of the full data set into $S$ equal sized partitions (with the potential exception of the final partition). These subsets are unique, meaning there is no overlap (of observations) between partition $i$ and $j$ for every $i,j = 1, \dots , S$. \\
We then fit the \gls{pim} on each partition and obtain $\widehat{\beta}^*_s$ estimates with $s = 1, \dots, S$. Calculating a final estimate for $\beta$ is straightforward via
\begin{align} \label{subsample_est}
\hat{\beta} = \frac{1}{S} \sum_{s = 1}^S \widehat{\beta}_{s}^*,
\end{align} 
where $\widehat{\beta}_{s}^*$ is the $s^{th}$ estimate for $\beta$ based on $\frac{n}{S}$ \gls{iid} data points. \\
We explore two ways to estimate $\Var{(\beta)}$, the variability of our final \gls{pim} estimator. The first one ignores the sandwich estimator for the variance of each $\widehat{\beta}_{s}^*$ and could naively be calculated as
\begin{align} \label{biased_var}
\widehat{\Var}{(\widehat\beta)} = \frac{1}{S - 1} \sum_{s = 1}^S \left(\widehat\beta^*_s - \overbar{\widehat\beta}^* \right)^2.
\end{align}
Importantly, equation (\ref{biased_var}) will lead to a biased estimation as the variability of any estimator obtained on a subset will differ from its variability on the full data set \citep{Kleiner2014, Wang2016}. However, as suggested in \cite{Politis1994}, one can scale the standard error (se) obtained on a subset to match its counterpart on the full data set. To see, let us take $X_1, \dots, X_n$ independent and random draws from a population with normal distribution $N(\mu_X, \sigma^2_X)$. It is known that $\Var{(\bar{X})} = \frac{1}{n} \Var{(X)}$. Hence if we consider the $S$ estimates $\widehat{\beta}^*$ as realizations from an independent sampling process, we have 
\begin{align} \label{scaled}
\text{se}(\widehat\beta) \approx \text{se}(\widehat{\beta}^*) \times \sqrt{\frac{1}{S}}.
\end{align}
The same rationale is used in the \textit{m out of n} bootstrap \citep{Bickel1997} where $m < n$ and one scales with $\sqrt{\frac{m}{n}}$ \citep{Geyer2013}. \\
We thus propose to scale $\widehat{\Var}{(\widehat\beta)}$ and construct a $1 - \alpha^*$ \gls{ci}, with $\alpha^*$ being the level of uncertainty through
\begin{align} \label{scaled_CI}
\hat{\beta} \pm z_{1 - \sfrac{\alpha^*}{2}} \times \text{se}(\widehat{\beta}^*) \times \sqrt{\frac{1}{S}}.
\end{align}
Here, $z$ is the quantile corresponding to the normal distribution and $\text{se}(\hat{\beta}^*)$ is the biased standard error of the estimator obtained by taking the square root of equation (\ref{biased_var}). We shall further refer to this \gls{ci} as the \textit{scaled confidence interval}, using the \textit{scaled standard error}. \\
Our second approach to estimate $\Var{(\beta)}$ is based on the $S$ sandwich variance estimates. Using equation (\ref{subsample_est}), we get:
\begin{align} \label{san_var} \nonumber
\widehat{\Var}{(\widehat\beta)} &= \Var{\left( \frac{1}{S} \sum_{s = 1}^S \hat\beta^*_s \right)} \\ \nonumber
 &= \frac{1}{S^2}\Var{\left( \sum_{s = 1}^S \hat\beta^*_s \right)} \\ 
 &= \frac{1}{S^2} \sum_{s = 1}^S \Var{\left(\hat\beta^*_s \right)} ,
\end{align}
where we substitute $\Var{\left(\hat\beta^*_s \right)}$ with the $s^{\text{th}}$ sandwich estimator from the \gls{pim} theory \citep{Thas2012}. Note, equation (\ref{san_var}) is correct, assuming there is no dependence between \textit{partitions}. We construct a second \gls{ci} through:
\begin{align} \label{san_var_CI}
\hat{\beta} \pm z_{1 - \sfrac{\alpha^*}{2}} \times \sqrt{ \frac{1}{S^2} \sum_{s = 1}^S \Var{\left(\hat\beta^*_s \right)} }
\end{align}
and refer to this as the \textit{adjusted sandwich estimator \gls{ci}} using the \textit{adjusted standard error}.

\subsection{Uniform subsampling}
For our second algorithm, we introduce the set of subsampling probabilities $\pi_i, i = 1,...,n$ assigned to all data points and $\pi = \{\pi_i\}_{i=1}^n$. We define $K$ as the number of subsampled observations with $K \ll n$ and $B$ as the amount of resampling iterations. \\

First, we assign uniform subsampling probabilities to all data points. That is $\pi^{\text{UNI}} = \{\pi_i = n^{-1}\}_{i=1}^n$. Then we start the first iteration in which $K$ observations are sampled without replacement from the original data set. We then fit the \gls{pim} on this subset, save the estimates and iterate the procedure for $B$ times. Note that different iterations are not independent of each other. This approach is summarized in box 1.\\
We obtain the estimate for $\beta$ using equation (\ref{subsample_est}) while replacing subscripts $(S;s)$ with $(B;b)$. The variance of the estimator is again first calculated as the \textit{scaled standard error}. We now have $\frac{n}{K}$ equally sized parts. Hence
\begin{align} 
\text{se}(\widehat\beta) \approx \text{se}(\widehat{\beta}^*) \times \sqrt{\frac{K}{n}}. 
\end{align}
Thus we construct a $1 - \alpha^*$ \gls{ci} through
\begin{align} \label{scaled_ci_uniform_sampling}
\hat{\beta} \pm z_{1 - \sfrac{\alpha^*}{2}} \times \text{se}(\widehat{\beta}^*) \times \sqrt{\frac{K}{n}}.
\end{align}
Second, we replace subscripts in equation (\ref{san_var}) for the \textit{adjusted standard error} and obtain its corresponding $1 - \alpha^*$ \gls{ci} through
\begin{align}  \label{san_var_uni}
\widehat{\Var}{(\widehat\beta)}  &= \frac{1}{B^2} \sum_{b = 1}^B \Var{\left(\hat\beta^*_b \right)}.
\end{align}
Crucially, recall that we make the assumption of the $B$ iterations being independent. However, this is not the case here as we might sample duplicate observations between iterations. Especially if $B$ increases, then $\widehat{\Var}{(\widehat\beta)}$ will be underestimated as equation (\ref{san_var_uni}) ignores the covariance between iterations. \\
For this reason, we hypothesize that it will be better to use the \textit{scaled standard error} for this algorithm. \\

In the following section, we shall discuss the different data generating models used in our simulation study.

\begin{figure}[!t]
\begin{tcolorbox}[colback=green!5,colframe=green!40!black,title=Box 1: uniform subsampling algorithm]
Define $\boldsymbol{\pi}^{\text{UNI}} = \{\pi_i = n^{-1}\}_{i=1}^n$. Then the uniform subsampling algorithm proceeds in four steps:

1) \textbf{sample}: draw a random sample $K \ll n$ without replacement from the original dataset with probability $\boldsymbol{\pi}^{\text{UNI}}$ \\
2) \textbf{estimate}: estimate the $\beta$ parameters of a \gls{pim} on the subset of the data. \\
3) \textbf{iterate}: repeat step 1 and 2 for \textit{B} times. \\
4) \textbf{aggregate} the $B$ estimates.

\end{tcolorbox}
\end{figure}

\section{Data generating models}
First, we establish a relationship between \gls{pim} and univariate linear regression models. The latter are used to generate data for the Monte Carlo simulations. Generally, it is true that multiple parametric models can be used to generated data under a semiparametric model \citep{Thas2012}. However, \cite{Thas2012} show that if the linear regression model holds, there is a proportional relationship between estimated \gls{pim} parameters (i.e. $\hat{\beta}$) and those estimated using \gls{ols} in linear regression models ($\hat{\alpha}$). The asymptotic properties of the \gls{pim} estimators also hold under different data generating models \citep{Thas2012}. Though for compactness, we restrict our data generating models to linear regression models and leave futher explorations under different settings for follow-up research. \\
Thereafter we will discuss the three data generating models to test the performance of the sampling algorithms. Each Monte Carlo simulation generates a $n = 250.000$ data set. \\

\subsection{Relationship between $\beta$ and $\alpha$}
To demonstrate the relationship between \gls{pim} and linear regression models \citep{Thas2012}, consider the following linear model with a one-dimensional covariate $X$:
\begin{align} \label{lm}
Y = \mu + \alpha X + \varepsilon,
\end{align}
where $\varepsilon$ is normally distributed with mean 0 and variance $\sigma^2$.

Now have a continuous $Y$ which allows the \gls{pi} to take the form:
\begin{align} \label{PIMoflm} \nonumber
P(Y \preceq Y^{*}|X, X^{*}) &= P(Y < Y^{*}|X, X^{*}) \\ \nonumber
&= P(\mu + \alpha X + \varepsilon < \mu + \alpha X^{*} + \varepsilon^{*} | X, X^{*}) \\ \nonumber
&= P(\varepsilon - \varepsilon^{*}< \mu - \mu + \alpha X^{*} - \alpha X | X, X^{*}) \\ \nonumber
&=P\{\varepsilon - \varepsilon^{*} < \alpha (X^{*} - X) | X, X^{*} \} \\ 
&= F_{\Delta}\{\alpha (X^{*} - X)\},
\end{align}
where $F_{\Delta}$ is the cumulative distribution function of $\varepsilon - \varepsilon^{*}$. 
Then, if we know that all observations are sampled independently and both $\varepsilon \stackrel{d}{=} N(0, \sigma^{2})$ and $\varepsilon^{*} \stackrel{d}{=} N(0, \sigma^{2})$, it is true that $F_{\Delta}$ is also normally distributed with mean 0 and variance equal to
\begin{align} \label{var} \nonumber
\Var{(\varepsilon - \varepsilon^{*})} &= \Var{(\varepsilon)} + \Var{(\varepsilon^{*})} - 2 \times \Cov{(\varepsilon, \varepsilon^{*})} \\ \nonumber
 &= \Var{(\varepsilon)} + \Var{(\varepsilon^{*})} \\ \nonumber
 &= \sigma^{2} + \sigma^{2} \\ 
 &= 2\sigma^{2}.
\end{align}

The corresponding \gls{pim} for equation (\ref{PIMoflm}) with a link function $g(.) = F^{-1}_{\Delta}(.)$ equals
\begin{align} \label{link}
g\{P(Y < Y^{*}|X, X^{*})\} &= g[F_{\Delta}\{\alpha (X^{*} - X)\}].
\end{align}

By suggesting the probit link function and combining the results from equation (\ref{var}) and (\ref{link}), there is
\begin{align} \label{lm_pim} \nonumber
g\{P(Y < Y^{*}|X, X^{*})\} &= g \left [\Phi \left \{ \frac{\alpha (X^{*} - X)}{\sqrt{2\sigma^{2}}} \right \} \right ] \\
&= \frac{\alpha (X^{*} - X)}{\sqrt{2} \sigma}.
\end{align}

Finally, consider the \gls{pim} as described in equation (\ref{Chap2:eqPIM}) for a continuous $Y$ and a one-dimensional covariate $X$. Using the usual linear predictor $Z = X^{*} - X$ and a probit link function, \cite{Thas2012} establish the relationship between linear regression models and \gls{pim} models through $\beta = \alpha / \sqrt{2}\sigma$. \\

We shall now discuss three linear models that are used to generate data. 

\subsection{Model 1} \label{data_model1}
For the first model, we generate data under the following linear model:
\begin{align*} 
Y_i = \alpha X_i + \varepsilon_i
\end{align*}

with $i = 1, \dots, n$ and $\varepsilon_i$ are \gls{iid} $N(0, \sigma^2)$.  The predictor $X$ is uniformly sampled from the interval $[0.1, u]$, where $u$ = 1. Furthermore, we set $\sigma^2 = 1$ and $\alpha = 5$. \\
The true value of $\beta$ in this model equals $\frac{5}{\sqrt{2}} = 3.54$.

\subsection{Model 2} \label{data_model2}
The structural form of the second model is equal to model 1. We now generate data with $\alpha =1$, $u =10$ and $\sigma = 5$. The true value of the \gls{pim} estimator eqals $\frac{1}{\sqrt{2 \times 25}} = 0.14$.

\subsection{Model 3} \label{data_model3}
The third model is a multiple regression model in which we attempt to replicate the data set observed in \cite{Przybylski2017} as we will analyse their research question using a \gls{pim} in Chapter \ref{Chapter:real_data}. Without providing too much detail here, the authors of this paper used linear regression to model the effect (among other variables) of smartphone usage in the weekdays (measured on a Likert scale from 0-7; $X$) on self-reported mental well being. To control for confounding effects, they included variables for sex (female = 0, male = 1; $Z_{1}$), whether the participant was located in economic deprived areas (no = 0, yes = 1; $Z_{2}$) and the ethnic background (minority group no = 0, yes = 1; $Z_{3}$) in the model. The full model is given as:
\begin{align*} 
Y_i = \mu + \alpha X_i + \gamma_1 Z_{1i} + \gamma_2 Z_{2i} + \gamma_3 Z_{31i} + \varepsilon_i 
\end{align*} 

Based on a linear regression on the full observed dataset (where $n = 116.630$, complete cases), we find the regression parameters, the proportions of the covariates and the standard deviation of the outcome variable ($\sigma = 9.51$). These parameters are then used to generate $n = 250.000$ observations where $\varepsilon_i$ are again \gls{iid} $N(0, \sigma^2)$. See Table \ref{table:LM_coef} for an overview of these parameters. Note that we still sample the predictor uniformly, though now from the interval $[0, 7]$ where $X$ is restricted to integers. \\

\begin{table}[!h] \centering 
\captionsetup{width=0.9\textwidth}
\begin{tabular}{lcc} 
\toprule
\multicolumn{1}{c}{Parameter} & \multicolumn{1}{c}{Estimate} & \multicolumn{1}{c}{Average (proportion of occurence)} \\
\midrule
Intercept ($\mu$) & 46.717 &  // \\ 
smartphone usage ($\alpha$) & $-$0.432 & // \\ 
sex (F = 0, M = 1, $\gamma_1$) & 4.550 & 0.48 \\ 
minority (no = 0, yes = 1, $\gamma_2$) & 0.305 & 0.24 \\ 
deprived (no = 0, yes = 1, $\gamma_3$) & $-$0.451 & 0.43 \\ 
\bottomrule
\end{tabular} 
  \caption{Estimated data generating parameters for model 3 and the proportion of each covariate taking the value 1. Parameters estimated using \gls{ols} on the full dataset of \cite{Przybylski2017}.} 
  \label{table:LM_coef} 
\end{table} 

\begin{figure}[!ht]
\begin{center}
\includegraphics[width = 0.80\textwidth]{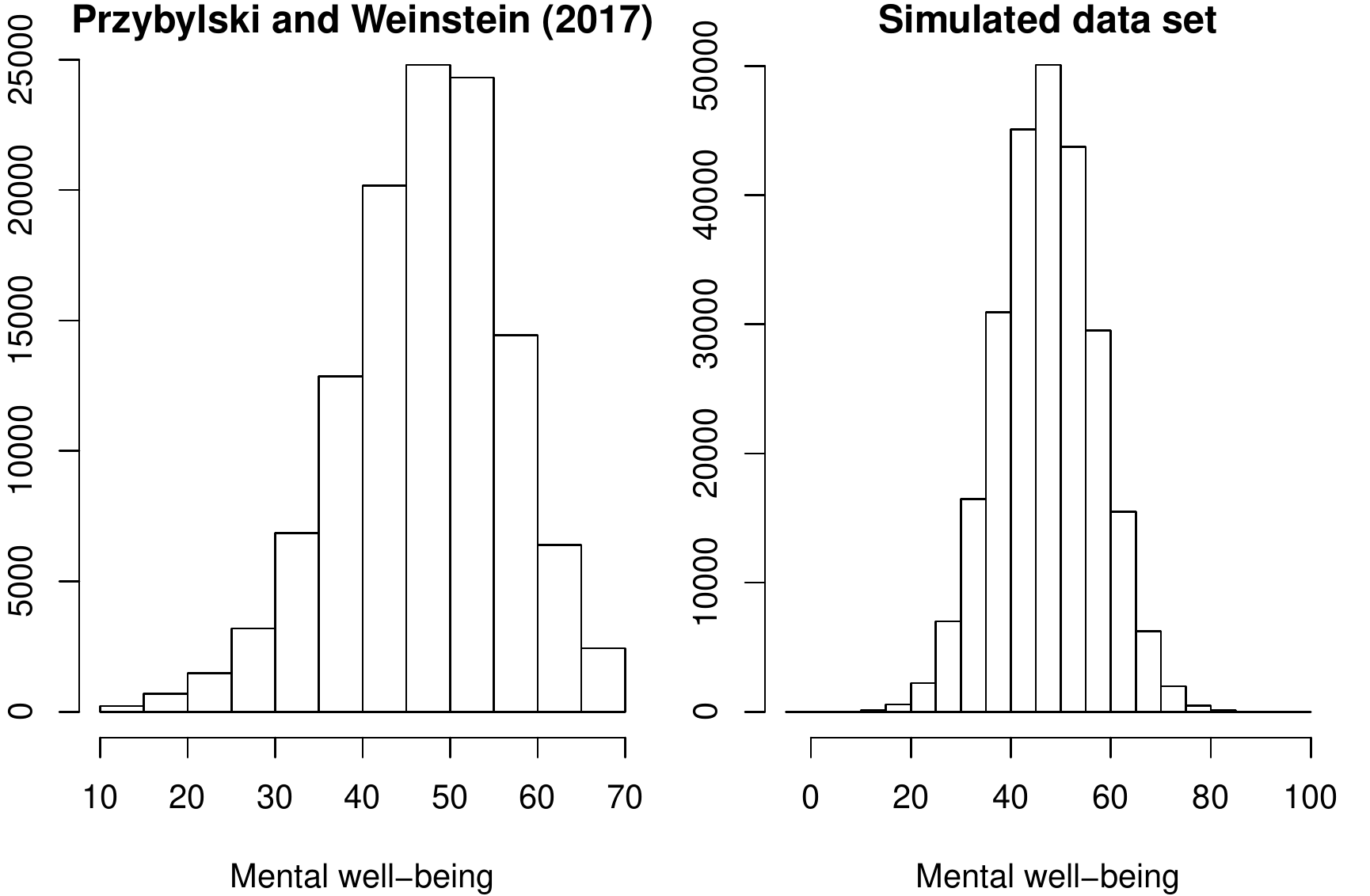}
\caption{Histogram of the observed mental well being (left) in \cite{Przybylski2017} and an example of one generated dataset (right) for model 3.}\label{hist_MWB}
\end{center}
\end{figure}

An example of one simulated dataset compared with the actual dataset is depicted in Figure \ref{hist_MWB}. Note the slight amount of skewness in the original dataset which we did not model in the simulations. \\
The true value of the \gls{pim} estimator for $\beta$ in this model equals $\frac{-0.43}{\sqrt{2} \times 9.51} = -0.0319722$.

\section{Simulation study}
To evaluate the different choices of \textit{K} and \textit{B} with respect to the \textit{uniform subsampling} algorithm, we create a grid of $10 \times 10$ combinations in which we let both \textit{K} and \textit{B} range from 10 to 1000. Thus we have $K \text{ and } B \in \{10, 120, 230, 340, 450, 560, 670, 780, 890, 1000\}$. For each combination between those parameters, we have 1000 Monte Carlo runs which leads to $10 \times 10 \times 1000$ simulations. As mentioned above, we generate $n = 250.000$ observations per simulation. \\
For the \textit{single data partitioning}, we consider $S = 100$ \textit{partitions} of each $K = 2500$ data points.  \\
Both algorithms share the same simulated data at the start of each Monte Carlo simulation run. This is done to ensure a fair comparison between the two procedures. \\

We are interested in the average time to fit a \gls{pim} as well as key properties of the observed distribution of the \gls{pim} estimators over all Monte Carlo runs. These include (1) bias quantified using the mean squared error, (2) the normal approximation of the estimators through plotting $\hat{\beta}$ from each simulation run in QQ-plots. And (3) the empirical coverage of the 95\% confidence intervals around $\hat\beta$ using both the \textit{scaled} and the \textit{adjusted standard error}. \\

All simulations are performed using \code{R} \citep{RCoreTeam2013} and High Performance Computing (HPC). We estimate the \gls{pim} parameters using the \textit{pim} package \citep{pimPackage2017} available on CRAN. \\
Note that although we run the simulations using HPC, we will consider the time to fit a \gls{pim} on a single machine in Chapter \ref{Chapter:real_data}.

	\chapter{Results of simulation study} \label{Chapter:SimRes}

\section{Single data partitioning}
\begin{figure}[!h]
\begin{center}
\includegraphics[scale = 0.6]{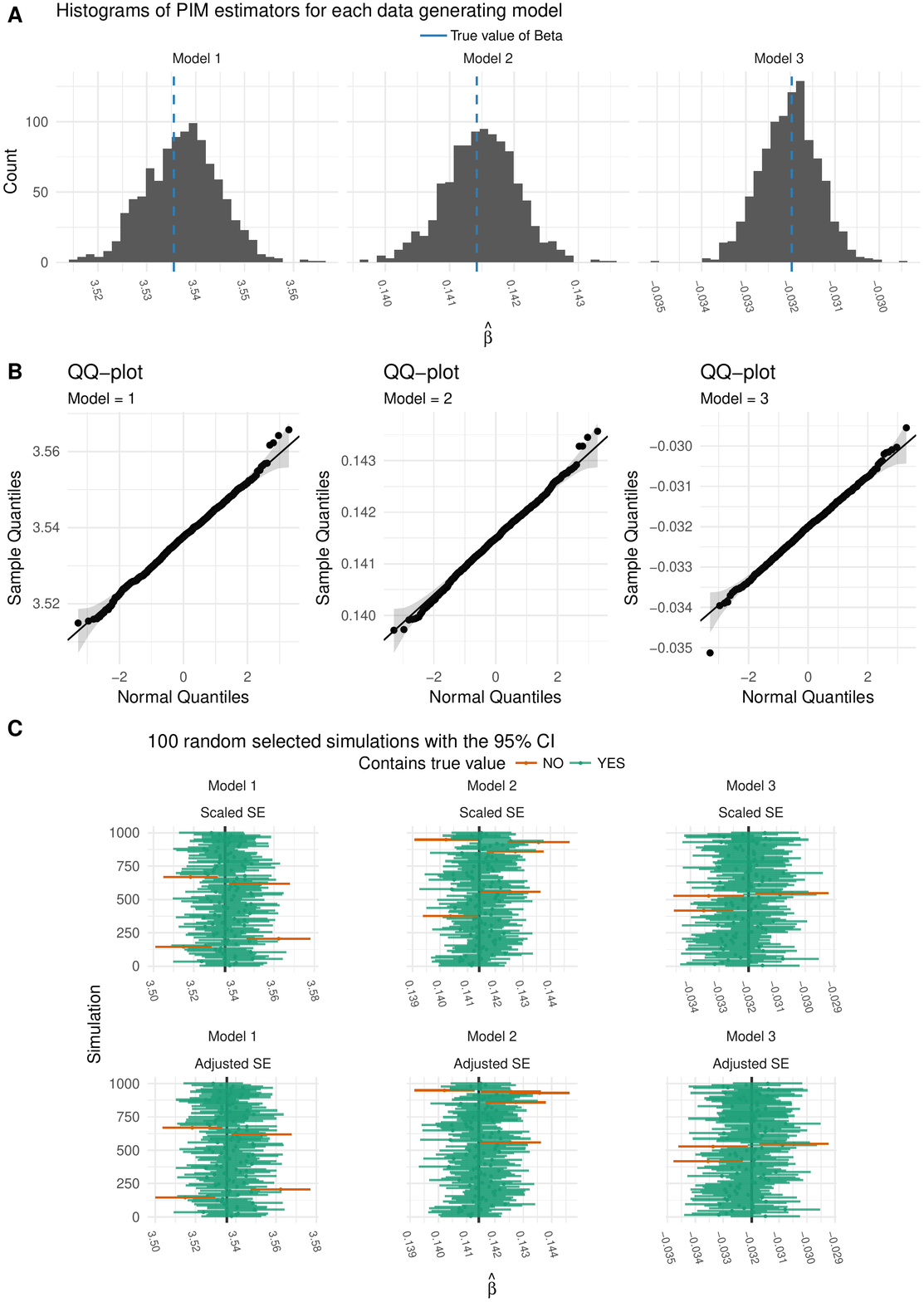}
\caption{Simulation results based on 1000 Monte Carlo runs for the single data partitioning algorithm under all 3 data generating models. A: histogram of the $\beta$ \gls{pim} estimates over all simulations with the vertical line being the true parameter. B: normal QQ-plots. C: 100 randomly selected simulations showing the 95\% \textit{scaled \gls{ci}} and the \textit{adjusted sandwich estimator \gls{ci}} for $\beta$, provided as a reference.} \label{key_results_partition}
\end{center}
\end{figure}

Simulation results based on 1000 Monte Carlo runs using the \textit{single data partitioning} algorithm are shown in Table \ref{table:part_results} and Figure \ref{key_results_partition}. First, we do not observe large differences between the three data generating models from Section \ref{data_model1}, \ref{data_model2} and \ref{data_model3}. \\
With respect to the variance of the estimator, we also observe small differences between the \textit{scaled} (eq. (\ref{scaled})) and \textit{adjusted} (eq. (\ref{san_var})) calculations. The empirical coverages of the 95\% confidence intervals for $\beta$ are close to the nominal level of $1 - \alpha^{*} = 0.95$ using both standard error calculations. Furthermore, the sample variance of the simulated $\hat\beta$ is nearly identical as the average (over all simulations) of the variance \gls{pim} estimates.  \\
Next, histograms of $\hat\beta$ (eq. (\ref{subsample_est})) in Figure \ref{key_results_partition} are centered around the true parameter $\beta$, with the exception of the histogram under the first data generating model. There are some deviations in the tails of the normal QQ-plots, although the normal approximation is reasonable. \\

Note, due to the partitioning of the data, one can reduce the time to fit a \gls{pim} by distributing computations to several nodes. As we use high performance computing, it is not sensible to record the time it takes to fit a model (ideally, one also takes into account the time it takes to transfer data). For this reason, we shall provide estimates of time needed to fit a \gls{pim} in Chapter \ref{Chapter:real_data} where we use a single machine only. \\ 
In general, the \textit{single data partitioning} algorithm under current settings is associated with favourable results.


\begin{table}[!ht] \centering 
\captionsetup{width=0.9\textwidth}
\begin{tabular}{lccccc} 
\toprule
\multicolumn{1}{c}{Model} & \multicolumn{1}{c}{MSE} & \multicolumn{1}{c}{$\Var{\left(\hat\beta\right)}$} & \multicolumn{1}{c}{Type of \gls{ci}} & \multicolumn{1}{c}{EC} & \multicolumn{1}{c}{$\text{Av}\left(\hat{S}^{2}_{\hat\beta}\right)$} \\
\midrule
\multirow{2}{*}{1} & \multirow{2}{*}{5.9877e-05} & \multirow{2}{*}{5.6683e-05} & \cellcolor{black!20} scaled & \cellcolor{black!20} 0.952 & \cellcolor{black!20} 5.7981e-05\\
&  & & ASE & 0.946 & 5.7704e-05\\
\multirow{2}{*}{2} & \multirow{2}{*}{3.2321e-07} & \multirow{2}{*}{3.1919e-07} & \cellcolor{black!20} scaled & \cellcolor{black!20} 0.933 & \cellcolor{black!20} 2.9814e-07\\
&  & & ASE & 0.936 & 2.9651e-07\\
\multirow{2}{*}{3} & \multirow{2}{*}{4.3536e-07} & \multirow{2}{*}{4.3293e-07} & \cellcolor{black!20} scaled & \cellcolor{black!20} 0.944 & \cellcolor{black!20} 4.0226e-07\\
&  & & ASE & 0.946 & 4.0245e-07\\
\bottomrule
\end{tabular} 
  \caption{Simulation results based on 1000 Monte Carlo runs for the single data partitioning algorithm. Shown in columns are the mean squared error (MSE), $\Var{\left(\hat\beta\right)}$ is the sample variance of the simulated $\hat\beta$, the empirical coverage (EC) of the 95\% \gls{ci} for $\beta$ and the average of the variance \gls{pim}  estimates. The latter is generically denoted as $\text{Av}\left(\hat{S}^{2}_{\hat\beta}\right)$, but either calculated and used for the \textit{scaled \gls{ci}} (eq. (\ref{scaled_CI})) or the \textit{adjusted sandwich estimator \gls{ci}} (ASE, eq. (\ref{san_var_CI})). All 3 data generating models are given row wise.}
  \label{table:part_results} 
\end{table} 

\section{Uniform subsampling}
\begin{figure}[!h]
\begin{center}
\includegraphics[width = 0.90\textwidth]{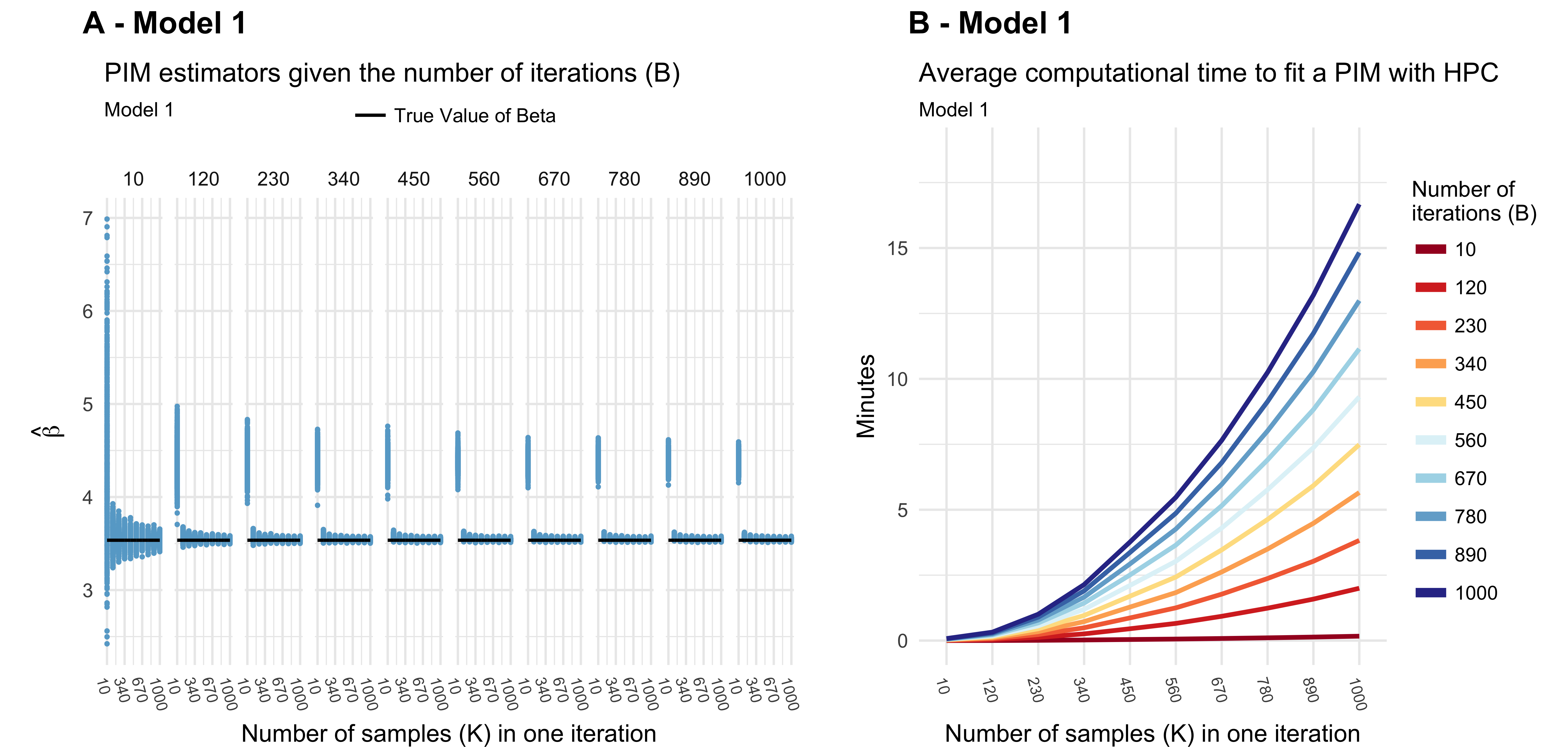}
\caption{Simulation results for 1000 Monte Carlo runs using the uniform sampling algorithm. Data is generated under model 1 (Section \ref{data_model1}). A: $\beta$ \gls{pim} estimates with respect to the true parameter for all $K$ and $B$. B: average time to fit a \gls{pim}.} \label{key_results_uni_model_1}
\end{center}
\end{figure}

\begin{figure}[!h]
\begin{center}
\includegraphics[width = 0.90\textwidth]{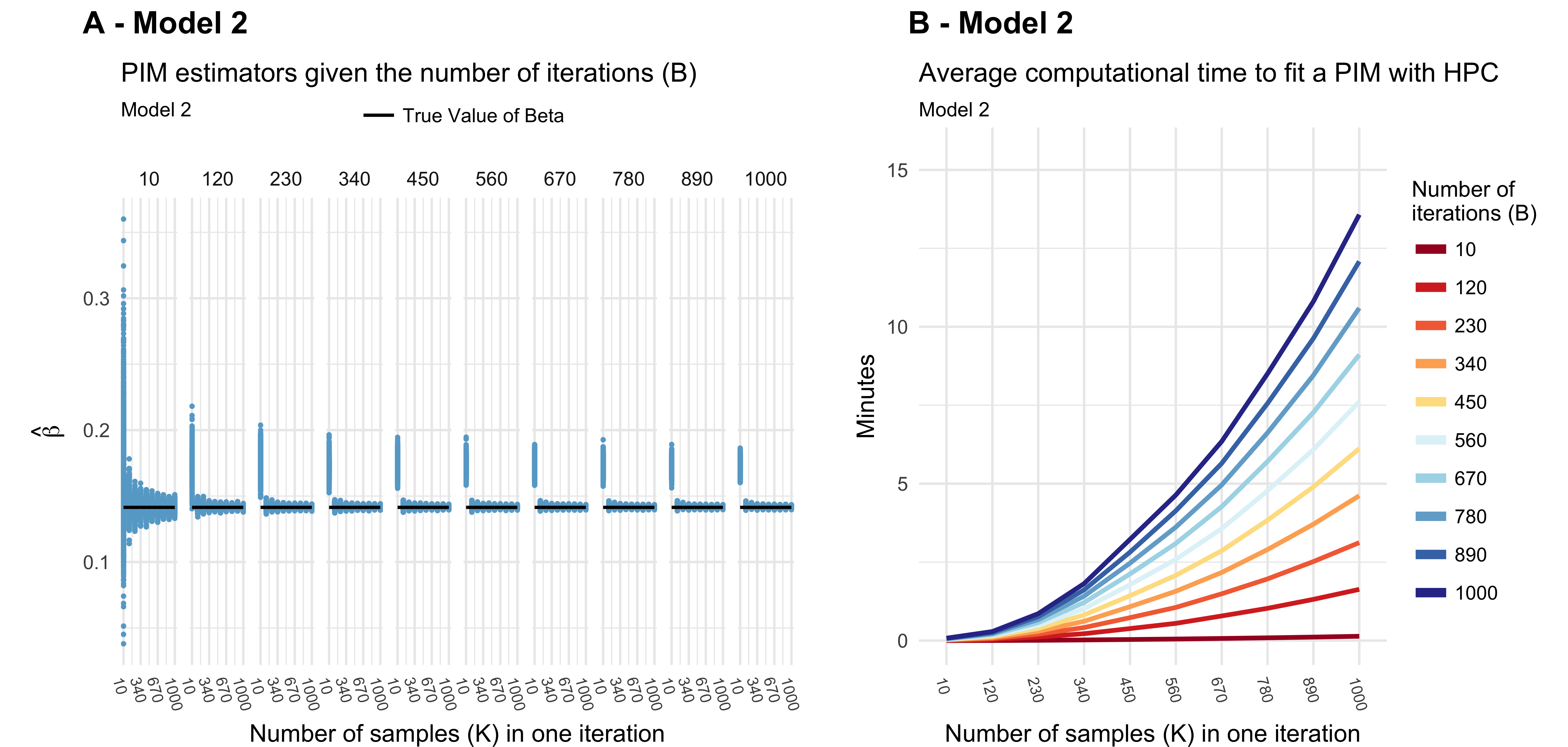}
\caption{Simulation results for 1000 Monte Carlo runs using the uniform sampling algorithm. Data is generated under model 2 (Section \ref{data_model2}). A: $\beta$ \gls{pim} estimates with respect to the true parameter for all $K$ and $B$. B: average time to fit a \gls{pim}.} \label{key_results_uni_model_2}
\end{center}
\end{figure}

\begin{figure}[!h]
\begin{center}
\includegraphics[width = 0.90\textwidth]{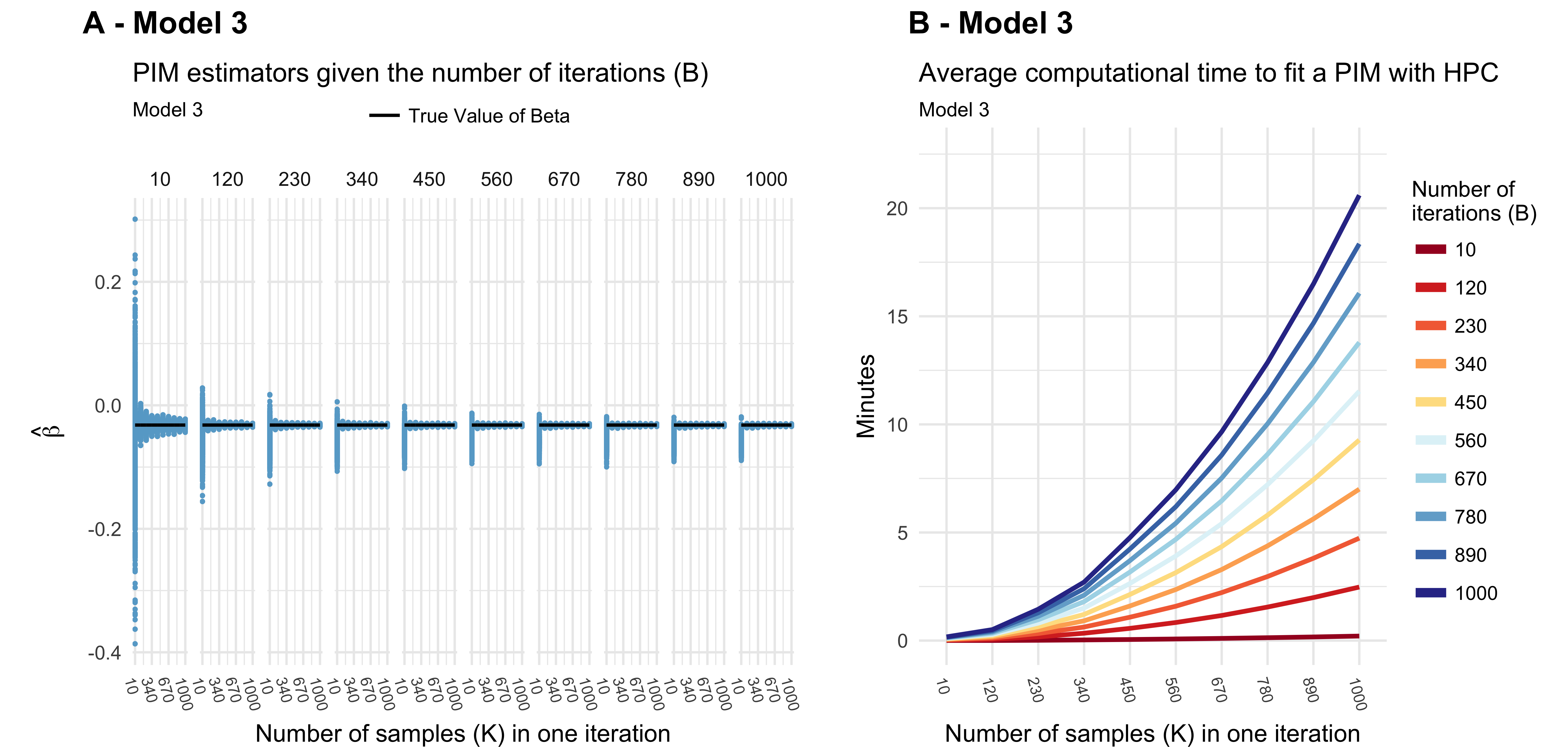}
\caption{Simulation results for 1000 Monte Carlo runs using the uniform sampling algorithm. Data is generated under model 3 (Section \ref{data_model3}). A: $\beta$ \gls{pim} estimates with respect to the true parameter for all $K$ and $B$. B: average time to fit a \gls{pim}.} \label{key_results_uni_model_3}
\end{center}
\end{figure}

Selected simulation results using the \textit{uniform subsampling} algorithm are shown in Figure \ref{key_results_uni_model_1}, \ref{key_results_uni_model_2} and \ref{key_results_uni_model_3} for the data generating models from Section \ref{data_model1}, \ref{data_model2} and \ref{data_model3} respectively. These figures contain the $\beta$ \gls{pim} estimates with respect to the true parameter and the average time to fit a \gls{pim}. Note that we have restricted each execution of a \textit{uniform subsampling} algorithm to one computation node. This allows to record the time to fit a \gls{pim}. The mean squared errors are given in Tables \ref{MSE_12} and \ref{MSE_3}. The empirical coverages of the 95\% confidence intervals for $\beta$ using both the \textit{scaled} as well as the \textit{adjusted standard error} calculation are given in Table \ref{EC_uniform_m1}, \ref{EC_uniform_m2} and \ref{EC_uniform_m3}. We will discuss all results irrespective of the data generating model, as the patterns between them are fairly similar. \\

To start with, $\beta$ seems consistently estimated using $\hat\beta$ as the estimates are converging to the true value of $\beta$ when both $K$ and $B$ increase (see panel A of Figure \ref{key_results_uni_model_1}, \ref{key_results_uni_model_2} and \ref{key_results_uni_model_3}). Furthermore, we observe estimates that are reasonable close to the true value as soon as both the number of subsampled observations and the amount of resampling iterations equal 230. This is important as the average time to estimate the \gls{pim} parameters using these settings equals $19.7\ (\text{sd} = 0.007)$ seconds for data generating model 3. Using $K = 450$ and $B = 450$ already results in an average of $2.13\ (\text{sd} = 0.02)$ minutes. \\

\begin{figure}[!h]
\begin{center}
\includegraphics[width = 0.90\textwidth]{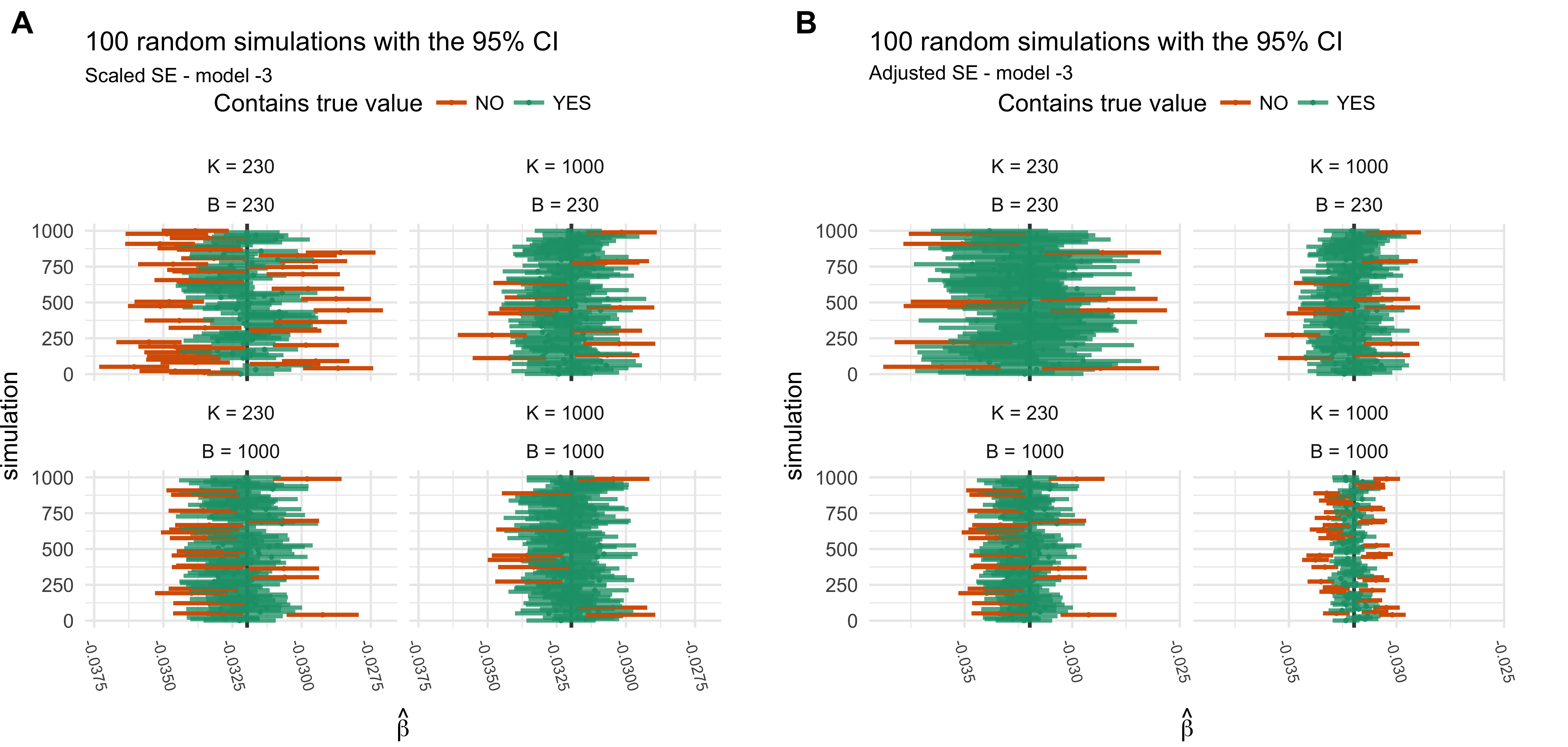}
\caption{Simulation results for 1000 Monte Carlo runs using the uniform sampling algorithm. Data is generated under model 3 (Section \ref{data_model3}). A: 100 randomly selected simulations showing the 95\% \textit{scaled \gls{ci}} for $\beta$ when K = 230 and B = 1000. B: the same 100 randomly selected simulations showing the 95\% \textit{adjusted sandwich estimator \gls{ci}} for $\beta$.} \label{CI_results_model_3}
\end{center}
\end{figure}

In contrast with the \textit{single data partitioning} algorithm, we do find differences between the calculation of the \textit{scaled} versus the \textit{adjusted standard error}. We refer to Table \ref{EC_uniform_m3} to formulate four main findings with respect to the empirical coverages of the 95\% \gls{ci}s for $\beta$ under data generating model 3 (Section \ref{data_model3}). Same results hold for the other two data generating models. \\
First, empirical coverages using the \textit{scaled standard error} are (slowly) approaching $0.95$ as both $K$ and $B$ increase. On the contrary, this is not the case for the \textit{adjusted sandwich estimator \gls{ci}}. As $B$ increases, the performance gets worse. Furthermore, the length of the \gls{ci} seems to decrease using the \textit{adjusted sandwich estimator} as $B$ increases. See for instance Figure \ref{CI_results_model_3}. Finally, the observed coverages are generally low, unless we use the \textit{adjusted sandwich estimator \gls{ci}} with $K = 1000$ and $B = 10$. \\

\begin{figure}[!h]
\begin{center}
\includegraphics[width = 0.85\textwidth]{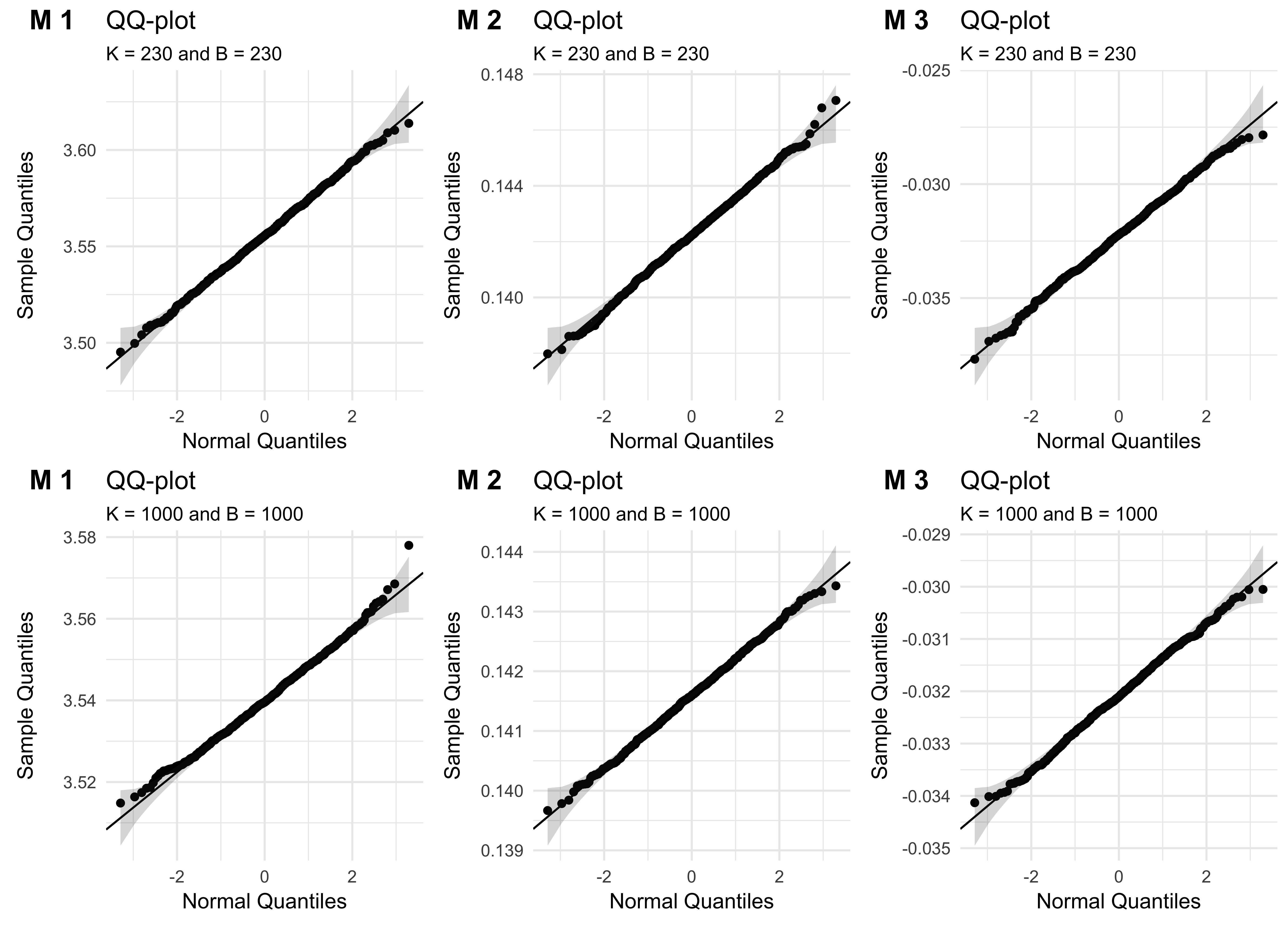}
\caption{Normal QQ-plots of $\hat\beta$ for both $K$ and $B$ = 230 and $K$ and $B$ = 1000. Each column corresponds to a data generating model (left model 1, middle model 2 and right model 3).} \label{QQ_results_uni}
\end{center}
\end{figure}

Figure \ref{QQ_results_uni} shows the normal QQ-plots for the combinations between $K = \{230, 1000\}$ and $B = \{230, 1000\}$ under all three data generating models. In general, we observe a reasonable normal approximation, although there are slight deviations from normality in the tails. Normal QQ-plots for all other combinations of $K$ and $B$ are similar and can be found in the appendix (Figure \ref{appendix:QQ_mod1}, \ref{appendix:QQ_mod2} and \ref{appendix:QQ_mod3}). \\

Finally, when comparing the mean squared error of the \textit{single data partitioning} (Table \ref{table:part_results}) with the \textit{uniform subsampling} (Table \ref{MSE_12} and \ref{MSE_3}), we observe values of the same magnitude. Especially when $K$ and $B$ are sufficiently large. Moreover, we also compare the relative variance of the \gls{pim} estimates of the \textit{uniform subsampling} for $K = B = 1000$ over the variance of the \textit{single data partitioning}. We find for data generating model 1, 2 and 3 the relative variance being equal to  $1.28$, $1.19$ and $1.15$ respectively. This indicates that the estimates obtained through the \textit{uniform subsampling} algorithm have a slightly higher variability.

\afterpage{%
 \clearpage
 \thispagestyle{empty}
\begin{landscape}
\centering 
\begin{tabular}{lllllllllll}
\multicolumn{1}{c}{} & \multicolumn{10}{c}{MSE MODEL 1} \\ 
\toprule
  & \textbf{B = 10} & \textbf{B = 120} & \textbf{B = 230} & \textbf{B = 340} & \textbf{B = 450} & \textbf{B = 560} & \textbf{B = 670} & \textbf{B = 780} & \textbf{B = 890} & \textbf{B = 1000}\\
\midrule
\rowcolor{black!20} \textbf{K = 10} & 1.2155e+00 & 7.6160e-01 & 7.4476e-01 & 7.3521e-01 & 7.2808e-01 & 7.2188e-01 & 7.1961e-01 & 7.1735e-01 & 7.1270e-01 & 7.1048e-01\\
\rowcolor{black!20} \textbf{K = 120} & 1.5066e-02 & 2.9115e-03 & 2.3595e-03 & 2.1759e-03 & 2.0534e-03 & 1.9846e-03 & 1.9494e-03 & 1.9144e-03 & 1.8901e-03 & 1.8665e-03\\
\rowcolor{black!20} \textbf{K = 230} & 6.7774e-03 & 1.0084e-03 & 7.5580e-04 & 6.4252e-04 & 6.1059e-04 & 5.7761e-04 & 5.6251e-04 & 5.5288e-04 & 5.4399e-04 & 5.3689e-04\\
\rowcolor{black!20} \textbf{K = 340} & 4.4479e-03 & 6.5625e-04 & 4.5246e-04 & 3.8325e-04 & 3.4652e-04 & 3.1659e-04 & 3.1128e-04 & 3.0442e-04 & 2.9800e-04 & 2.9715e-04\\
\rowcolor{black!20} \rowcolor{black!20} \textbf{K = 450} & 3.2917e-03 & 4.4949e-04 & 3.0896e-04 & 2.6586e-04 & 2.3910e-04 & 2.2401e-04 & 2.1817e-04 & 2.1248e-04 & 2.0782e-04 & 2.0524e-04\\

\textbf{K = 560} & 2.7479e-03 & 3.4936e-04 & 2.5335e-04 & 2.1763e-04 & 1.9028e-04 & 1.7454e-04 & 1.6793e-04 & 1.6229e-04 & 1.5813e-04 & 1.5613e-04\\
 \textbf{K = 670} & 2.2307e-03 & 2.7285e-04 & 2.0765e-04 & 1.7153e-04 & 1.5585e-04 & 1.4789e-04 & 1.4177e-04 & 1.3791e-04 & 1.3529e-04 & 1.3181e-04\\
\textbf{K = 780} & 1.9095e-03 & 2.6379e-04 & 1.8250e-04 & 1.5691e-04 & 1.3969e-04 & 1.3544e-04 & 1.2758e-04 & 1.2418e-04 & 1.2050e-04 & 1.2002e-04\\
 \textbf{K = 890} & 1.7163e-03 & 2.2334e-04 & 1.6119e-04 & 1.3828e-04 & 1.2343e-04 & 1.1498e-04 & 1.1033e-04 & 1.0808e-04 & 1.0628e-04 & 1.0249e-04\\
\textbf{K = 1000} & 1.6339e-03 & 2.0026e-04 & 1.4426e-04 & 1.1780e-04 & 1.1093e-04 & 1.0465e-04 & 9.9593e-05 & 9.6505e-05 & 9.3790e-05 & 9.1918e-05\\

\multicolumn{1}{c}{} & \multicolumn{10}{c}{MSE MODEL 2} \\ 
\midrule
\rowcolor{black!20} \textbf{K = 10} & 2.8661e-03 & 1.1504e-03 & 1.0975e-03 & 1.0447e-03 & 1.0452e-03 & 1.0334e-03 & 1.0264e-03 & 1.0227e-03 & 1.0144e-03 & 1.0101e-03\\
\rowcolor{black!20} \textbf{K = 120} & 7.2636e-05 & 7.7936e-06 & 5.6566e-06 & 4.6031e-06 & 4.3590e-06 & 4.0492e-06 & 3.8370e-06 & 3.6584e-06 & 3.5203e-06 & 3.4475e-06\\
 \rowcolor{black!20}\textbf{K = 230} & 3.4441e-05 & 3.8067e-06 & 2.4646e-06 & 1.9984e-06 & 1.7546e-06 & 1.6242e-06 & 1.4859e-06 & 1.4054e-06 & 1.3549e-06 & 1.3290e-06\\
\rowcolor{black!20}\textbf{K = 340} & 2.2980e-05 & 2.4947e-06 & 1.5595e-06 & 1.2508e-06 & 1.1182e-06 & 1.0188e-06 & 9.6394e-07 & 9.1449e-07 & 8.6588e-07 & 8.4914e-07\\
\rowcolor{black!20} \textbf{K = 450} & 1.6479e-05 & 1.7353e-06 & 1.1424e-06 & 9.3196e-07 & 8.3624e-07 & 7.6371e-07 & 6.9456e-07 & 6.5529e-07 & 6.4295e-07 & 6.2739e-07\\

\textbf{K = 560} & 1.3737e-05 & 1.6617e-06 & 1.0253e-06 & 8.5070e-07 & 6.9957e-07 & 6.5100e-07 & 6.2589e-07 & 5.8504e-07 & 5.7678e-07 & 5.4210e-07\\
\textbf{K = 670} & 1.1517e-05 & 1.2655e-06 & 8.3668e-07 & 7.0117e-07 & 6.3215e-07 & 5.8377e-07 & 5.5177e-07 & 5.2464e-07 & 5.1244e-07 & 5.0277e-07\\
\textbf{K = 780} & 9.7427e-06 & 1.1497e-06 & 7.6779e-07 & 6.8207e-07 & 6.0953e-07 & 5.6886e-07 & 5.2946e-07 & 5.1559e-07 & 4.9158e-07 & 4.7730e-07\\
 \textbf{K = 890} & 9.1643e-06 & 1.0422e-06 & 7.3367e-07 & 6.0760e-07 & 5.6116e-07 & 5.2255e-07 & 4.8976e-07 & 4.6526e-07 & 4.5467e-07 & 4.4807e-07\\
\textbf{K = 1000} & 7.9593e-06 & 9.6029e-07 & 6.5872e-07 & 5.6144e-07 & 5.0649e-07 & 4.8399e-07 & 4.5371e-07 & 4.3493e-07 & 4.2148e-07 & 4.0717e-07\\

\bottomrule
\end{tabular}
        \captionof{table}{Mean squared error (MSE) of estimates under data generating model 1 and 2 using uniform subsampling for all levels of $K$ and $B$.} \label{MSE_12}
    \end{landscape}
    \clearpage
}

\afterpage{%
 \clearpage
 \thispagestyle{empty}
\begin{landscape}
\centering 
\begin{tabular}{lllllllllll}
\multicolumn{1}{c}{} & \multicolumn{10}{c}{MSE MODEL 3} \\ 
\toprule
  & \textbf{B = 10} & \textbf{B = 120} & \textbf{B = 230} & \textbf{B = 340} & \textbf{B = 450} & \textbf{B = 560} & \textbf{B = 670} & \textbf{B = 780} & \textbf{B = 890} & \textbf{B = 1000}\\
\midrule
\rowcolor{black!20}\textbf{K = 10} & 9.3666e-03 & 1.3358e-03 & 1.0234e-03 & 8.9582e-04 & 8.3303e-04 & 7.8536e-04 & 7.5128e-04 & 7.2819e-04 & 7.3131e-04 & 7.2198e-04\\
\rowcolor{black!20}\textbf{K = 120} & 8.8719e-05 & 8.7812e-06 & 5.0587e-06 & 3.6961e-06 & 3.0344e-06 & 2.6873e-06 & 2.4180e-06 & 2.2345e-06 & 2.1281e-06 & 1.9959e-06\\
\rowcolor{black!20}\textbf{K = 230} & 4.7386e-05 & 4.4971e-06 & 2.5413e-06 & 1.9086e-06 & 1.5586e-06 & 1.3374e-06 & 1.1834e-06 & 1.0892e-06 & 1.0305e-06 & 9.9710e-07\\
\rowcolor{black!20}\textbf{K = 340} & 3.1201e-05 & 2.7131e-06 & 1.7941e-06 & 1.3921e-06 & 1.1679e-06 & 1.0284e-06 & 9.7347e-07 & 9.2330e-07 & 8.8298e-07 & 8.5331e-07\\
\rowcolor{black!20}\textbf{K = 450} & 2.4273e-05 & 2.4064e-06 & 1.5221e-06 & 1.1844e-06 & 1.0208e-06 & 9.2425e-07 & 8.6173e-07 & 8.1159e-07 & 7.6885e-07 & 7.4419e-07\\
\addlinespace
\textbf{K = 560} & 1.8164e-05 & 1.9711e-06 & 1.1833e-06 & 9.4972e-07 & 8.6487e-07 & 7.9861e-07 & 7.5390e-07 & 7.1285e-07 & 6.6975e-07 & 6.4248e-07\\
\textbf{K = 670} & 1.5445e-05 & 1.8050e-06 & 1.1537e-06 & 9.8545e-07 & 8.6572e-07 & 7.8470e-07 & 7.2616e-07 & 6.8193e-07 & 6.4907e-07 & 6.3739e-07\\
\textbf{K = 780} & 1.3615e-05 & 1.4628e-06 & 9.7627e-07 & 7.8653e-07 & 6.8920e-07 & 6.4597e-07 & 6.1457e-07 & 5.8999e-07 & 5.8364e-07 & 5.6953e-07\\
\textbf{K = 890} & 1.2294e-05 & 1.2294e-06 & 8.5759e-07 & 7.3498e-07 & 6.7831e-07 & 6.3184e-07 & 6.0593e-07 & 5.7533e-07 & 5.6335e-07 & 5.5044e-07\\
\textbf{K = 1000} & 1.0522e-05 & 1.2235e-06 & 8.3147e-07 & 7.2404e-07 & 6.4770e-07 & 5.8663e-07 & 5.5929e-07 & 5.4686e-07 & 5.3113e-07 & 5.1049e-07\\
\bottomrule
\end{tabular}
        \captionof{table}{Mean squared error (MSE) of estimates under data generating model 3 using uniform subsampling for all levels of $K$ and $B$.} \label{MSE_3}
    \end{landscape}
    \clearpage
}

\afterpage{%
 \clearpage
 \thispagestyle{empty}
\begin{landscape}
\centering 
\begin{tabular}{lllllllllll} 
\multicolumn{1}{c}{} & \multicolumn{10}{c}{SCALED CI - EC MODEL 1} \\ 
\toprule
  & \textbf{B = 10} & \textbf{B = 120} & \textbf{B = 230} & \textbf{B = 340} & \textbf{B = 450} & \textbf{B = 560} & \textbf{B = 670} & \textbf{B = 780} & \textbf{B = 890} & \textbf{B = 1000}\\
\midrule
\rowcolor{black!20} \textbf{K = 10} & 0.01 & 0.00 & 0.00 & 0.00 & 0.00 & 0.00 & 0.00 & 0.00 & 0.00 & 0.00\\
\rowcolor{black!20} \textbf{K = 120}  & 0.09 & 0.17 & 0.14 & 0.11 & 0.09 & 0.06 & 0.05 & 0.04 & 0.04 & 0.03\\
\rowcolor{black!20} \textbf{K = 230}  & 0.13 & 0.33 & 0.36 & 0.35 & 0.35 & 0.34 & 0.34 & 0.33 & 0.31 & 0.31\\
\rowcolor{black!20} \textbf{K = 340}  & 0.18 & 0.41 & 0.47 & 0.51 & 0.53 & 0.54 & 0.54 & 0.54 & 0.54 & 0.53\\
\rowcolor{black!20} \textbf{K = 450}  & 0.22 & 0.51 & 0.59 & 0.62 & 0.64 & 0.66 & 0.68 & 0.68 & 0.68 & 0.68\\

\textbf{K = 560} & 0.23 & 0.55 & 0.63 & 0.69 & 0.72 & 0.73 & 0.75 & 0.76 & 0.76 & 0.77\\
\textbf{K = 670} & 0.24 & 0.62 & 0.69 & 0.74 & 0.76 & 0.78 & 0.79 & 0.80 & 0.81 & 0.81\\
\textbf{K = 780} & 0.25 & 0.64 & 0.74 & 0.77 & 0.79 & 0.80 & 0.81 & 0.82 & 0.82 & 0.83\\
\textbf{K = 890} & 0.27 & 0.66 & 0.78 & 0.80 & 0.82 & 0.83 & 0.85 & 0.85 & 0.86 & 0.87\\
\textbf{K = 1000} & 0.27 & 0.69 & 0.80 & 0.84 & 0.86 & 0.86 & 0.87 & 0.88 & 0.89 & 0.89\\
\multicolumn{1}{c}{} & \multicolumn{10}{c}{ADJUSTED SANDWICH ESTIMATOR CI - EC MODEL 1} \\ 
\rowcolor{black!20}\textbf{K = 10} & 0.46 & 0.00 & 0.00 & 0.00 & 0.00 & 0.00 & 0.00 & 0.00 & 0.00 & 0.00\\
\rowcolor{black!20}\textbf{K = 120} & 0.92 & 0.69 & 0.53 & 0.40 & 0.30 & 0.23 & 0.18 & 0.14 & 0.11 & 0.08\\
\rowcolor{black!20}\textbf{K = 230} & 0.93 & 0.83 & 0.73 & 0.66 & 0.57 & 0.49 & 0.44 & 0.39 & 0.34 & 0.32\\
\rowcolor{black!20}\textbf{K = 340} & 0.94 & 0.85 & 0.77 & 0.70 & 0.65 & 0.62 & 0.55 & 0.51 & 0.47 & 0.42\\
\rowcolor{black!20}\textbf{K = 450} & 0.95 & 0.88 & 0.81 & 0.74 & 0.70 & 0.65 & 0.62 & 0.57 & 0.54 & 0.51\\

\textbf{K = 560} & 0.94 & 0.88 & 0.82 & 0.75 & 0.72 & 0.67 & 0.66 & 0.61 & 0.59 & 0.56\\
\textbf{K = 670} & 0.94 & 0.89 & 0.81 & 0.76 & 0.72 & 0.67 & 0.64 & 0.62 & 0.59 & 0.56\\
\textbf{K = 780} & 0.94 & 0.86 & 0.80 & 0.75 & 0.72 & 0.67 & 0.64 & 0.61 & 0.58 & 0.57\\
\textbf{K = 890} & 0.95 & 0.86 & 0.81 & 0.74 & 0.70 & 0.66 & 0.65 & 0.62 & 0.58 & 0.57\\
\textbf{K = 1000} & 0.94 & 0.88 & 0.82 & 0.76 & 0.71 & 0.68 & 0.64 & 0.61 & 0.58 & 0.56\\
\bottomrule
\end{tabular}
        \captionof{table}{Empirical coverages of the 95\% \textit{scaled} and \textit{adjusted sandwich estimator \gls{ci}} for $\beta$ under data generating model 1 using \textit{uniform subsampling}.} \label{EC_uniform_m1}
    \end{landscape}
    \clearpage
}

\afterpage{%
 \clearpage
 \thispagestyle{empty}
\begin{landscape}
\centering 
\begin{tabular}{lllllllllll}
\multicolumn{1}{c}{} & \multicolumn{10}{c}{SCALED CI - EC MODEL 2} \\ 
\toprule
  & \textbf{B = 10} & \textbf{B = 120} & \textbf{B = 230} & \textbf{B = 340} & \textbf{B = 450} & \textbf{B = 560} & \textbf{B = 670} & \textbf{B = 780} & \textbf{B = 890} & \textbf{B = 1000}\\
\midrule
\rowcolor{black!20} \textbf{K = 10} & 0.02 & 0.00 & 0.00 & 0.00 & 0.00 & 0.00 & 0.00 & 0.00 & 0.00 & 0.00\\
\rowcolor{black!20} \textbf{K = 120} & 0.10 & 0.28 & 0.31 & 0.35 & 0.33 & 0.34 & 0.32 & 0.33 & 0.33 & 0.33\\
\rowcolor{black!20} \textbf{K = 230} & 0.14 & 0.42 & 0.50 & 0.52 & 0.55 & 0.59 & 0.59 & 0.61 & 0.62 & 0.63\\
\rowcolor{black!20} \textbf{K = 340} & 0.15 & 0.50 & 0.62 & 0.65 & 0.69 & 0.72 & 0.73 & 0.73 & 0.74 & 0.74\\
\rowcolor{black!20} \textbf{K = 450} & 0.20 & 0.58 & 0.68 & 0.73 & 0.76 & 0.78 & 0.79 & 0.81 & 0.82 & 0.81\\

\textbf{K = 560} & 0.20 & 0.57 & 0.72 & 0.74 & 0.80 & 0.82 & 0.84 & 0.84 & 0.85 & 0.86\\
\textbf{K = 670} & 0.24 & 0.67 & 0.76 & 0.80 & 0.82 & 0.84 & 0.86 & 0.86 & 0.87 & 0.88\\
\textbf{K = 780} & 0.25 & 0.68 & 0.78 & 0.81 & 0.83 & 0.84 & 0.87 & 0.87 & 0.88 & 0.87\\
\textbf{K = 890} & 0.28 & 0.70 & 0.79 & 0.83 & 0.84 & 0.85 & 0.88 & 0.88 & 0.88 & 0.88\\
\textbf{K = 1000} & 0.28 & 0.74 & 0.82 & 0.84 & 0.86 & 0.87 & 0.89 & 0.89 & 0.90 & 0.91\\
\multicolumn{1}{c}{} & \multicolumn{10}{c}{ADJUSTED SANDWICH ESTIMATOR CI - EC MODEL 2} \\ 
\rowcolor{black!20}\textbf{K = 10} & 0.68 & 0.08 & 0.01 & 0.00 & 0.00 & 0.00 & 0.00 & 0.00 & 0.00 & 0.00\\
\rowcolor{black!20}\textbf{K = 120} & 0.93 & 0.88 & 0.82 & 0.76 & 0.69 & 0.64 & 0.60 & 0.56 & 0.53 & 0.48\\
\rowcolor{black!20}\textbf{K = 230} & 0.95 & 0.91 & 0.86 & 0.82 & 0.78 & 0.75 & 0.73 & 0.69 & 0.66 & 0.64\\
\rowcolor{black!20}\textbf{K = 340} & 0.94 & 0.90 & 0.88 & 0.83 & 0.80 & 0.77 & 0.74 & 0.72 & 0.70 & 0.68\\
\rowcolor{black!20}\textbf{K = 450} & 0.95 & 0.92 & 0.88 & 0.85 & 0.80 & 0.78 & 0.74 & 0.73 & 0.71 & 0.68\\

\textbf{K = 560} & 0.95 & 0.89 & 0.85 & 0.82 & 0.80 & 0.77 & 0.74 & 0.72 & 0.69 & 0.67\\
\textbf{K = 670} & 0.94 & 0.91 & 0.86 & 0.82 & 0.78 & 0.74 & 0.73 & 0.69 & 0.67 & 0.65\\
\textbf{K = 780} & 0.95 & 0.88 & 0.85 & 0.79 & 0.74 & 0.72 & 0.68 & 0.66 & 0.65 & 0.62\\
\textbf{K = 890} & 0.94 & 0.89 & 0.84 & 0.78 & 0.74 & 0.72 & 0.68 & 0.65 & 0.64 & 0.61\\
\textbf{K = 1000} & 0.94 & 0.89 & 0.83 & 0.77 & 0.72 & 0.69 & 0.65 & 0.63 & 0.62 & 0.60\\
\bottomrule
\end{tabular}
        \captionof{table}{Empirical coverages of the 95\% \textit{scaled} and \textit{adjusted sandwich estimator \gls{ci}} for $\beta$ under data generating model 2 using \textit{uniform subsampling}.} \label{EC_uniform_m2}
    \end{landscape}
    \clearpage
}

\afterpage{%
 \clearpage
 \thispagestyle{empty}
\begin{landscape}
\centering 
\begin{tabular}{lllllllllll}
\multicolumn{1}{c}{} & \multicolumn{10}{c}{SCALED CI - EC MODEL 3} \\ 
\toprule
  & \textbf{B = 10} & \textbf{B = 120} & \textbf{B = 230} & \textbf{B = 340} & \textbf{B = 450} & \textbf{B = 560} & \textbf{B = 670} & \textbf{B = 780} & \textbf{B = 890} & \textbf{B = 1000}\\
\midrule
\rowcolor{black!20}\textbf{K = 10} & 0.02 & 0.08 & 0.06 & 0.06 & 0.05 & 0.04 & 0.03 & 0.03 & 0.02 & 0.02\\
\rowcolor{black!20}\textbf{K = 120} & 0.12 & 0.32 & 0.44 & 0.50 & 0.56 & 0.58 & 0.60 & 0.61 & 0.62 & 0.62\\
\rowcolor{black!20}\textbf{K = 230} & 0.15 & 0.45 & 0.56 & 0.65 & 0.68 & 0.74 & 0.74 & 0.78 & 0.78 & 0.80\\
\rowcolor{black!20}\textbf{K = 340} & 0.16 & 0.55 & 0.65 & 0.71 & 0.75 & 0.78 & 0.79 & 0.80 & 0.82 & 0.82\\
\rowcolor{black!20}\textbf{K = 450} & 0.19 & 0.57 & 0.70 & 0.76 & 0.79 & 0.81 & 0.83 & 0.83 & 0.85 & 0.85\\

\textbf{K = 560} & 0.24 & 0.62 & 0.75 & 0.81 & 0.82 & 0.85 & 0.86 & 0.86 & 0.87 & 0.88\\
\textbf{K = 670} & 0.24 & 0.65 & 0.76 & 0.79 & 0.82 & 0.84 & 0.86 & 0.86 & 0.88 & 0.87\\
\textbf{K = 780} & 0.26 & 0.71 & 0.79 & 0.83 & 0.87 & 0.88 & 0.88 & 0.89 & 0.90 & 0.90\\
\textbf{K = 890} & 0.26 & 0.72 & 0.83 & 0.84 & 0.88 & 0.89 & 0.90 & 0.91 & 0.92 & 0.92\\
\textbf{K = 1000} & 0.31 & 0.74 & 0.83 & 0.85 & 0.88 & 0.89 & 0.90 & 0.90 & 0.91 & 0.92\\
\multicolumn{1}{c}{} & \multicolumn{10}{c}{ADJUSTED SANDWICH ESTIMATOR CI - EC MODEL 3} \\ 
\rowcolor{black!20}\textbf{K = 10} & 0.64 & 0 & 0 & 0 & 0 & 0 & 0 & 0 & 0 & 0\\
\rowcolor{black!20}\textbf{K = 120} & 0.94 & 0.93 & 0.91 & 0.90 & 0.87 & 0.85 & 0.84 & 0.82 & 0.80 & 0.79\\
\rowcolor{black!20}\textbf{K = 230} & 0.94 & 0.93 & 0.92 & 0.90 & 0.89 & 0.87 & 0.87 & 0.84 & 0.82 & 0.81\\
\rowcolor{black!20}\textbf{K = 340} & 0.95 & 0.94 & 0.91 & 0.88 & 0.85 & 0.85 & 0.82 & 0.78 & 0.77 & 0.76\\
\rowcolor{black!20}\textbf{K = 450} & 0.94 & 0.92 & 0.88 & 0.86 & 0.82 & 0.80 & 0.78 & 0.75 & 0.74 & 0.72\\

\textbf{K = 560} & 0.95 & 0.92 & 0.89 & 0.85 & 0.82 & 0.80 & 0.77 & 0.74 & 0.72 & 0.71\\
\textbf{K = 670} & 0.95 & 0.91 & 0.86 & 0.82 & 0.78 & 0.75 & 0.73 & 0.71 & 0.68 & 0.67\\
\textbf{K = 780} & 0.95 & 0.91 & 0.85 & 0.82 & 0.78 & 0.77 & 0.74 & 0.70 & 0.67 & 0.64\\
\textbf{K = 890} & 0.94 & 0.92 & 0.87 & 0.81 & 0.77 & 0.72 & 0.69 & 0.67 & 0.64 & 0.61\\
\textbf{K = 1000} & 0.94 & 0.89 & 0.85 & 0.78 & 0.74 & 0.71 & 0.69 & 0.66 & 0.63 & 0.61\\
\bottomrule
\end{tabular}
        \captionof{table}{Empirical coverages of the 95\% \textit{scaled} and \textit{adjusted sandwich estimator \gls{ci}} for $\beta$ under data generating model 2 using \textit{uniform subsampling}.} \label{EC_uniform_m3}
    \end{landscape}
    \clearpage
}

	\chapter{Application} \label{Chapter:real_data}
In this chapter, we demonstrate the \textit{single data partitioning} algorithm in combination with the \textit{adjusted sandwich estimator \gls{ci}} to fit a \gls{pim} on a large data set. We use this algorithm as the results from our simulation study showed it is associated with good performances. We will first introduce the setting and data set. We will then provide an analysis using linear regression models. Finally, we will use the \textit{single data partitioning} algorithm to fit \gls{pim}s. \\

\section{The relationship between digital screen usage and mental well-being}
While technology and digital devices are continuously shaping the lives of human beings, there is a growing concern in the field of developmental psychology about the extended time that children and adolescents spent using these devices. It has been suggested that prolonged usage of digital devices might be negatively associated with social and mental well-being (though see \cite{Bell2015} for a critical review). From theory, the \textit{digital Goldilocks hypothesis} states that while moderate usage is not harmful or may even be advantageous, spending too much time in front of digital screens potentially interferes with alternative activities such as socializing, sports, studying, etc. \citep{Przybylski2017}.

\begin{figure}[!ht]
\begin{center}
\includegraphics[scale=0.3]{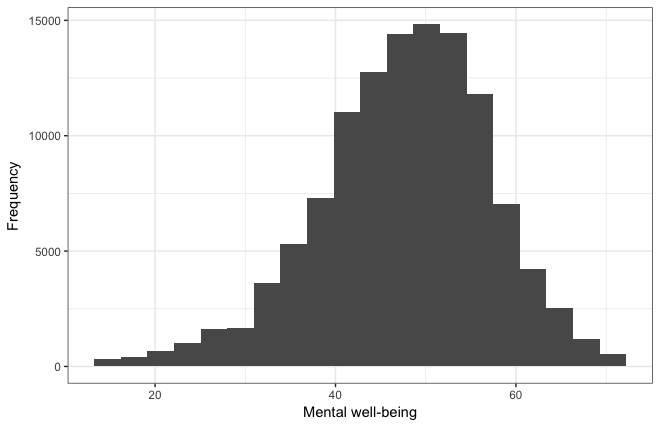}
\caption{Histogram of mental well-being.}\label{Chapter4:histogram}
\end{center}
\end{figure}

To investigate this hypothesis, \cite{Przybylski2017} conducted a large scale preregistered survey study in the United Kingdom (UK). While the original sampling framework contained $298,080$ participants, the final complete case data set contains $116,630$ 15-years old adolescents from the UK. The original analysis plan, code and data of this study are hosted on the Open Science Framework (\url{https://osf.io/82ybd/})\footnote{No common license provided}. \\
The participants were asked to fill in the Warwick-Edinburgh Mental Well-Being Scale \citep{Tennant2007}. This scale measures happiness, social well-being, psychological functioning and life satisfaction. It is a 14-item validated scale with a high internal consistency (Cronbachs $\alpha = .90$). The scores range from 14 to 70 (mean = $47.51$, SD = $9.51$) with high scores indicating higher mental well-being (\textit{MWBI}), see Figure \ref{Chapter4:histogram} for a histogram. \\

In the first part of their paper, \cite{Przybylski2017} have run 16 separate (univariate) analyses. These correspond to the combinations of 4 predictors measured either during the weekdays or weekends and fitted with or without control variables as covariates. The variables of interest are: watching movies, playing video games, computer usage and smartphone usage (\textit{SMART}). These were recorded using self-reported Likert scales, ranging from 0 to 7 hours of engagement. The first interval consists of half an hour.  The covariates are gender (\textit{GENDER}; female = 0, male = 1), whether the participant was located in economic deprived areas (\textit{DEPRIVED}; no = 0, yes = 1) and the ethnic background (\textit{MINORITY}; no = 0, yes = 1). Note that by running 16 separate univariate analyses, there is an inflation of the type I error rate. Indeed, it is advised to control for multiple testing such as a Bonferroni correction \citep{Bonferroni1936}. It should also be noted that the authors haver pre-registered the design and analysis plan of the study. With only small deviations from the analysis plan, there is less risk of distorted results due to \textit{p} hacking. An alternative approach would have been to go for a model building strategy with stepwise selection. \\

\begin{figure}[!ht]
\begin{center}
\includegraphics[scale=0.56]{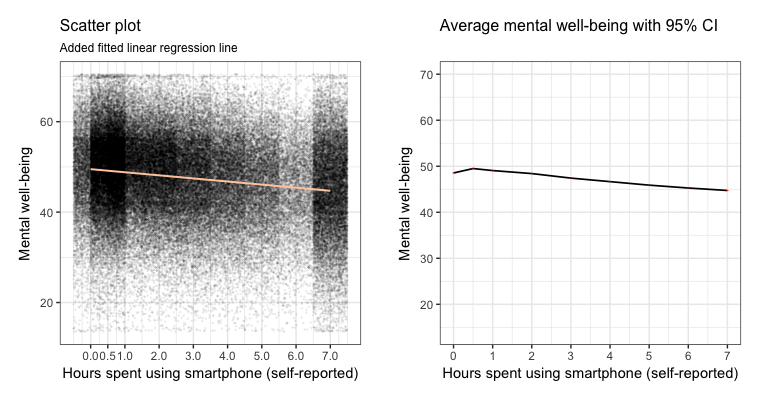}
\caption{Scatterplot (left) of mental well-being versus time spent using a smartphone during the weekdays. The solid line represents a fitted regression line with \gls{ols}. A small amount of horizontal jitter is added for visualization purpose. This jitter does not overlap into adjacent categories. The figure on the right is the average mental well-being versus time spent using a smartphone.}\label{scatter_average}
\end{center}
\end{figure}

For demonstration purpose, brevity and since the results for other predictors in the original paper are similar, we will focus only on the time spent using smartphones during the weekdays, once without covariates and once with covariates. \\
Before we do any analysis, we have a look at the data. A scatter plot of mental well-being versus the time spent with smartphones and the average mental well-being is given in Figure \ref{scatter_average}. Note the following two observations. First the distribution between the different levels of self-reported hours engaged with a smartphone is not uniform. There appears to be a ceiling effect as more observations are recorded in the last category (7 hours). Furthermore, the intervals are presumably perceived to be equally spaced by participants while this is not true for the interval $[0;1]$. This leads to a distorted visualization in the left part of the scatter plot. The second observation is the apparent curvilinear trend in the data (see the right part of Figure \ref{scatter_average}). We will address this trend in the second part of this chapter. In the first part, we consider the predictor as a continuous scale. We will model both the average response and the probabilistic index as a (monotonic) linear trend. \\

\subsection{Ordinary least squares}

\subsubsection{Linear trend}
For comparison, we begin with modelling the average mental well-being either without or with covariates:
\begin{align}
\E{(MWBI)} &= \alpha_0 + \alpha_1\text{ SMART} \label{E_lm_wc} \\
\E{(MWBI)} &= \alpha_0 + \alpha_1 \text{ SMART} + \alpha_2 \text{ GENDER} + \alpha_3 \text{ DEPRIVED} + \alpha_4 \text{ MINORITY} \label{E_lm_c}
\end{align}
To obtain the $\alpha$ estimates, we will fit linear regression models using \gls{ols}. These are given in the upper part of Table  \ref{table:app_results}. With \gls{ols}, the (linear) effect of adolescents spending time using their smartphones during the week on mental well-being is estimated as $-0.43 \times SMART$ (controlled for gender, wealth status and ethnic background). Consider for instance 15-year adolescents in the UK. On average, adolescents who spend 2, 4 or 6 hours time with their smartphone are associated with reporting a \textit{lower} average mental well-being of $0.86$, $1.72$ and $2.58$ points on the Warwick-Edinburgh Mental Well-Being Scale than adolescents who do not use their smartphone. The 95\% \gls{ci}s are given as $[-0.84; -0.88], [-1.70; -1.74] \text{ and } [-2.56; -2.60]$. 

\subsubsection{Curvilinear trend}
Although we do not use a formal goodness-of-fit test, it seems possible that a linear model is not appropriate for this data. The average mental well-being is characterized by a small increase before decreasing (see the right panel of Figure \ref{scatter_average}). By ignoring the curvilinear trend, we will might end up with wrong estimates. There exists other methods/models that will provide a better fit to this data, one of which being generalized additive models \citep{Hastie1990}. However, this is beyond the scope of this dissertation. Hence in this section we present one approach for dealing with this curvature using \gls{ols}. \\

\begin{figure}[!ht]
\begin{center}
\includegraphics[scale=0.55]{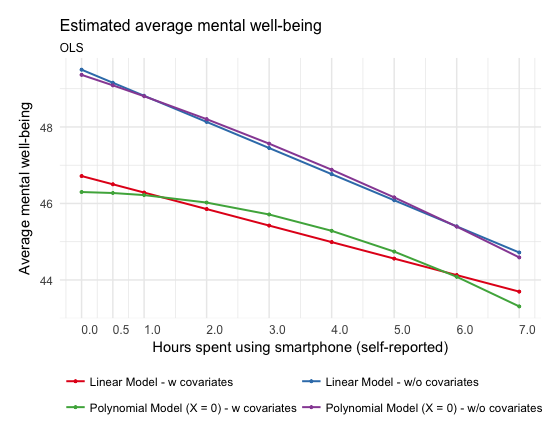}
\caption{Estimated average mental well-being versus time spent using a smartphone. In case of models with covariates: estimates correspond to girls from wealthy areas and belonging to an ethnic majority class.}\label{OLS_estimates}
\end{center}
\end{figure}

It is possible to model both a linear as well as a curvilinear trend by including a squared predictor in the model. For instance if we denote $X$ as smartphone usage, take the modelled mean response from equation (\ref{E_lm_wc}) and change it to:
\begin{align} \label{LM_squared}
\E{(MWBI)} &= \alpha_0 + \alpha_1 X + \alpha_2 X^{2} \ ,
\end{align}
then we have $\alpha_{2}$ estimating the curvature. Positive values indicate an upwards curvature, while negative values indicate a downwards curvature. Although $\alpha_1$ is associated with the linear component, its interpretation is not straightforward. On the other hand, if we take the first derivative of the right hand side of equation (\ref{LM_squared}), we get
\begin{align} 
\alpha_1 + 2 \alpha_2 X
\end{align}
This shows that $\alpha_{1}$ corresponds to the instantaneous rate of change, when $X$ = 0. In their paper \cite{Przybylski2017} have used equation (\ref{LM_squared}) and more particularly $\alpha_{2}$ to make a statement about the \textit{digital Goldilocks hypothesis}. Indeed, the authors observed a significant $\alpha_2 < 0$. However we would like to add two elements to their analysis. First, the authors ignored the estimates for $\alpha_1$. Recall that the \textit{digital Goldilocks hypothesis} claims initial beneficial effects of using digital devices. Hence, $\alpha_1$ should be positive at $X = 0$ to get a starting upwards trend. Eventually these will get dominated by the quadratic negative term. Second, it is possible to describe linear effects, using $\alpha_1$. Indeed, when $X = 0$ (i.e. no smartphone used), it is not meaningful to have a statement about the effect of using a smartphone on mental well-being. An alternative though, which we suggest here, is to do mean based centering of $X$ before fitting the model for a second time. This gives $\alpha_{1}$ as the rate of change on mental well-being for adolescents with an average smartphone usage. \\
Estimates for the models with a squared predictor without and with covariates for both the case of $X = 0$ and the mean based centering are given in the upper part of Table \ref{table:quad_results}. Note that we obtain the same estimates for the case of $X = 0$ as in \cite{Przybylski2017} in the model without covariates. For the model with covariates, we obtain a different $\alpha_1$. This is because the authors used the original Likert-scale variable (ranging from 1-9), without recoding the variable to the corresponding hours (only for this model). \\
For the model without covariates, we observe a negative significant value for $\alpha_2 = -0.0205\ (t = -9.86, df = 116627, p < 0.001)$. However, this is also true for $\alpha_1$ when $X = 0\ (\alpha_1 = -0.6577, t = -47.50, df = 116627, p < 0.001)$ and when $X$ takes the value of the average amount of smartphone usage $(\alpha_1 = -0.3692, t = -25.79, df = 116627, p < 0.001)$. Note that $\alpha_1$ is not significant when covariates are added and $X = 0$.  \\
Using the estimates from Table \ref{table:quad_results}, we can visualize the expected average mental well-being. In Figure \ref{OLS_estimates}, we plot the estimated average mental well-being against smartphone usage during the week using the models with only a linear predictor as well as the polynomial models ($X = 0$). For the models with covariates, we only plot the estimates for girls from wealthy areas and belonging to an ethnic majority class. We can observe a downwards trend, though no upwards trend is estimated. 

\subsubsection{Concluding remarks for OLS}
\begin{figure}[!ht]
\begin{center}
\includegraphics[scale=0.5]{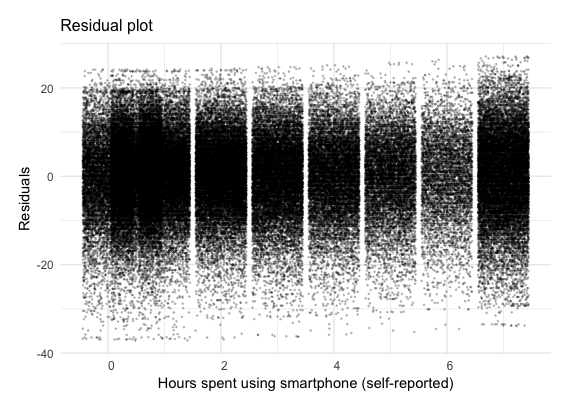}
\caption{Residuals plotted against the predictor. A small amount of horizontal jitter is added for visualization purposes. No jitter extends into adjecent categories. Residuals obtained by fitting the linear regression model with a squared predictor and covariates using \gls{ols}.}\label{residuals_OLS}
\end{center}
\end{figure}

\begin{figure}[!ht]
\begin{center}
\includegraphics[scale=0.5]{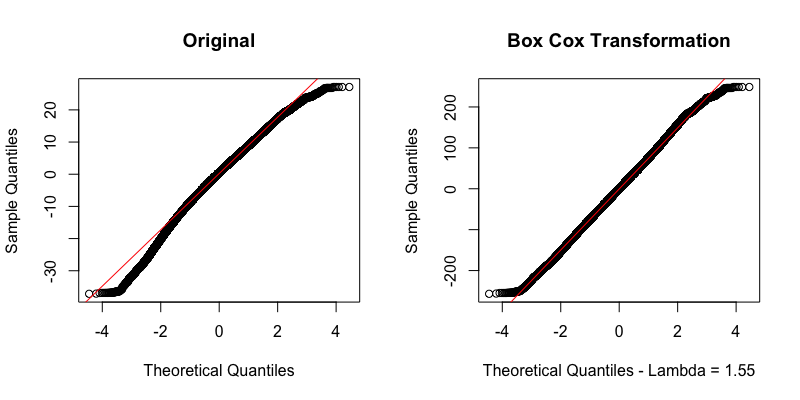}
\caption{Normal QQ plot of the residuals obtained by fitting the linear regression model with a squared predictor and covariates using \gls{ols}. Left: original response variable. Right: Box-Cox transformation on the response variable with $\lambda = 1.55$. }\label{QQ_OLS}
\end{center}
\end{figure}

So far, we have ignored the assumptions for the normal linear regression model. We briefly discuss these here before going to \gls{pim}s. We will focus only on model (\ref{LM_squared}) where we have included a squared predictor and covariates. To investigate the assumption of homoscedastic error terms, we plot the residuals versus the amount of time spent using a smartphone in Figure \ref{residuals_OLS}. Although not perfect, we consider no serious violation. Next, we also use the residuals in a normal QQ-plot (Figure \ref{QQ_OLS}) to investigate the normality assumption. Even though we have a large data set, there is quite a bit of deviation from normality in the tails. By using a Box-Cox transformation \citep{Box1982} on the outcome variable with $\lambda = 1.55$, we obtain a better approximation. Ideally, we should consider using the transformed response variable in our linear regression models. This seriously complicates the interpretation. However, an alternative would be to use \gls{pim}s. It is possible to show that if a transformation yields a reasonable approximation to normality, then using a \gls{pim} with the probit link function on the original variable is also valid. This is true due to the semiparametric nature of \gls{pim}s. \\
A second drawback of the analyses so far is that effects do not have a natural meaning as the response is measured on an ordinal scale. \\

For these reasons, we provide a different approach by modelling the \gls{pi}. 

\subsection{Probabilistic index models}
\subsubsection{Linear trend}

Consider the following probabilistic index models with probit link function:
\begin{align}
\text{probit} \{\text{P}(MWBI \preceq  MWBI^{*}) \} = &\beta_1(\text{SMART}^* - \text{SMART}) \label{pim_wc} \\ \nonumber
\text{probit}\{\text{P}(MWBI \preceq  MWBI^*) \} = & \beta_1(\text{SMART}^* - \text{SMART}) + \\ \nonumber
& \beta_2(\text{GENDER}^* - \text{GENDER}) + \\ \nonumber
& \beta_3(\text{DEPRIVED}^* - \text{DEPRIVED}) + \\
& \beta_4(\text{MINORITY}^* - \text{MINORITY}). \label{pim_c}
\end{align}

\begin{figure}[!ht]
\begin{center}
\includegraphics[scale=0.6]{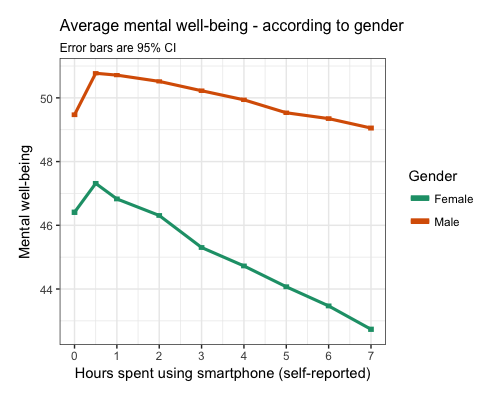}
\caption{Average mental well-being versus time spent using a smartphone for boys and girls.}\label{gender_difference}
\end{center}
\end{figure}

We use the \textit{single data partitioning} algorithm to fit these \gls{pim}s. With $116,630$ observations, we choose to split the data set at random into 117 partitions of each 1000 observations (with the last partition having 630 observations). The lower part of Table \ref{table:app_results} gives the model fits for the \gls{pim} estimators.  \\
Consider the comparison of two girls drawn at random from wealthy areas and without a minority background. The probability that girls who spent 0 hours on a smartphone during the week report a lower mental well-being compared to girls spending 2 hours, is estimated as $\Phi(\hat\beta_{1} \times \text{SMART}) = \Phi(-0.033 \times 2) = 47.37 \%$. Or vice versa, the probability of girls who spent 2 hours using their smartphones reporting a lower mental well-being compared to girls who spent 0 hours with their smartphone is estimated as $1 - 0.4737 = 52.63\%$. Likewise, the probabilities of girls who spent 4 or 6 hours reporting lower mental well-being compared to those with 0 hours are estimated as $55.25\%$ and $57.85\%$ respectively. The 95\% \gls{ci}s of those probabilities are $[52.55\%; 52.71\%], [55.17\%;  55.33\%] \text{ and } [57.77\%; 57.92\%]$. \\
Note how gender, the economic background and the ethnic background were only considered as covariates. Since these variables are not included in the original hypothesis, we did not include interaction terms into the models presented above. However if we ignore a potential interaction effect between gender and time using a smartphone on mental well-being, there is a profound difference between boys and girls. The probability of girls reporting a lower mental well-being compared to boys is equal to $64.06\%$ with a 95\% \gls{ci} of $[63.73\%;64.38\%]$. See Figure \ref{gender_difference} for an illustration. This plot also shows there is no clear suggestion of an interaction effects between gender with smartphone usage and mental well-being. \\

\begin{table}[!ht] \centering 
\captionsetup{width=0.9\textwidth}
\begin{tabular}{lcccc} 
\toprule
\multicolumn{1}{c}{Parameter} & \multicolumn{1}{c}{Estimate} & \multicolumn{1}{c}{Standard error} & \multicolumn{1}{c}{Test statistic} & \multicolumn{1}{c}{\textit{p} value} \\
\midrule
\multicolumn{5}{l}{\textit{Linear regression ordinary least squares - w/o covariates}} \\
Intercept $\left(\alpha_0\right)$ & $49.50$ & 0.0438 & $1129.06$ & $< 0.001$ \\
SMART $\left(\alpha_1\right)$ & $-0.68$ & 0.0117 & $-58.18$ & $< 0.001$ \\
\addlinespace
\multicolumn{5}{l}{\textit{Linear regression ordinary least squares - w covariates}} \\
Intercept $\left(\alpha_0\right)$ & $46.72$ & 0.0596 & $782.86$ & $< 0.001$ \\
SMART $\left(\alpha_1\right)$ & $-0.43$ & 0.0118 & $-36.48$ & $< 0.001$ \\
GENDER $\left(\alpha_2\right)$ & $4.55$ & 0.0551 & $82.52$ & $< 0.001$ \\
DEPRIVED $\left(\alpha_3\right)$ & $-0.45$ & 0.05578 & $4.68$ & $< 0.001$ \\
MINORITY $\left(\alpha_4\right)$ & $0.31$ & 0.0651 & $-8.08$ & $< 0.001$ \\
\addlinespace
\multicolumn{5}{l}{\textit{PIM - w/o covariates}} \\
SMART $\left(\beta_1\right)$ & $-0.052$ & 0.0009 & $-55.48$ & $<0.001$ \\
\addlinespace
\multicolumn{5}{l}{\textit{PIM - w covariates}} \\
SMART $\left(\beta_1\right)$ & $-0.033$ & 0.001 & $-34.29$ & $<0.001$ \\
GENDER $\left(\beta_2\right)$ & $0.36$ & 0.0044 & $81.85$ & $<0.001$ \\
DEPRIVED $\left(\beta_3\right)$ & $-0.036$ & 0.0044 & $-8.05$ & $<0.001$ \\
MINORITY $\left(\beta_4\right)$ & $0.017$ & 0.0052 & $3.36$ & $<0.001$ \\
\addlinespace
\bottomrule
\end{tabular} 
  \caption{Results of the ordinary least squares fits of the models without (w/o) and with (w) covariates and of the fits of the probabilistic index models without and with covariates. The test statistics correspond to $t$-values for \gls{ols} and $z$-statistics for \gls{pim}. Degrees of freedom for $t$-values in the model without covariates = 116628. With covariates = 116625.} \label{table:app_results} 
\end{table} 

A benefit of the \textit{single data partitioning} algorithm is the ability to run computations in parallel. As each partition is an independent subset of the original data set, we can run the partitioning once and distribute computations on each partition to the available computing resources. We have fitted model (\ref{pim_wc}) and (\ref{pim_c}) on a single machine, using 4 cores in parallel. The total time to fit each model and pool all estimates was equal to $0.3414$ and $0.5918$ minutes for the model without and with covariates respectively. \\
Finally, we have also fitted both of these models for a second time on a different data partitioning. This is done to check whether results are depending on an unknown predictor/covariate pattern which we are unable to capture due to the partitioning of the data set. We re-estimated $\beta_{1}$ for model (\ref{pim_wc}) as $-0.052\ (\text{SE} = 0.0009)$ and for model (\ref{pim_c}) as $-0.033\ (\text{SE} = 0.001)$, which are identical estimates. The time to estimate these two models was equal to 0.3310 and  0.5894 minutes. Hence we see that our results are stable across different partitioning outcomes. 

\subsubsection{Curvilinear trend}

\begin{figure}[!ht]
\begin{center}
\includegraphics[scale=0.6]{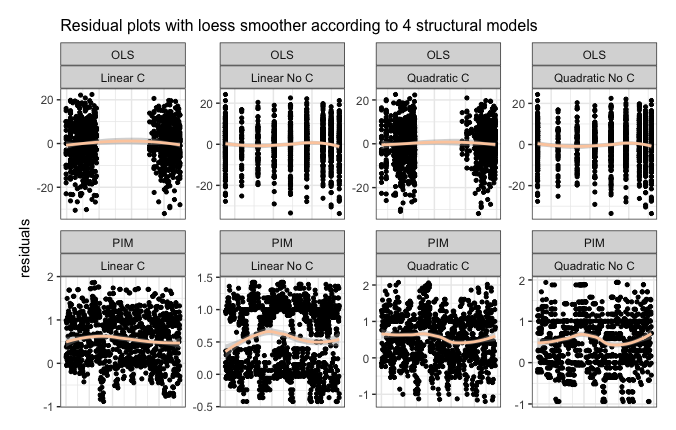}
\caption{Residual plots with a loess smoother. Top row: 1225 observations fitted with OLS, x-axis = predicted values. Bottom row: 1225 pseudo-observations with \gls{pim}, x-axis = index of pseudo-observation in the data set. Linear No C = linear model without covariates, Quadratic No C = + quadratic predictor, Linear C = linear model without covariates, Quadratic C = + quadratic predictor}\label{residual_plot}
\end{center}
\end{figure}

With \gls{pim} one is (generally) not interested in a linear effect in the first place. However, it can still be worthwhile to include a quadratic predictor in the model. Hence we have: 
\begin{align} \nonumber 
\text{probit} \{\text{P}(MWBI \preceq  MWBI^{*}) \} = &\beta_1(\text{SMART}^* - \text{SMART}) \\ \label{pim_noc_quad}
& \beta_2(\text{SMART}^{*2} - \text{SMART}^2)  
\end{align}
and a similar model with the covariates included. \\
We cannot visually see from Figure \ref{scatter_average} whether adding a squared predictor will improve the fit of our \gls{pim}. Furthermore, as is with any statistical model, the estimates can be wrong if the model, including the choice of the link function is not correctly specified. Unfortunately, assessing the goodness-of-fit for \gls{pim}s \citep{DeNeve2013a} on a large data set is no trivial task either as computation increases again with the amount of pseudo-observations. For now, we can get a rough idea by randomly selecting some observations and plot the residuals:
\begin{align}
R(X, X^*) = I(Y \preceq Y^*) - m(X,X^*;\hat\beta).
\end{align}
In Figure \ref{residual_plot}, we select 50 observations, which leads to 1225 pseudo-observations and plot the residuals versus their index in the data set with a LOESS (local regression) curve plotted. We do this for the linear \gls{pim} without and with covariates (model (\ref{pim_wc}) and (\ref{pim_c})) as well as the \gls{pim} with the quadratic predictor without and with covariates. We observe a more horizontal line when we restrict our analysis to a linear \gls{pim} with covariates. For a reference, we include residual plots for the same structural models, but fitted with \gls{ols}. We select 1225 original observations, fit the models and plot the residuals versus the predicted values. There is a small improvement when adding a quadratic predictor in the model, though the fitted line is not completely horizontal.

\begin{table}[!ht] \centering 
\captionsetup{width=0.9\textwidth}
\begin{tabular}{lcccc} 
\toprule
\multicolumn{1}{c}{Parameter} & \multicolumn{1}{c}{Estimate} & \multicolumn{1}{c}{Standard error} & \multicolumn{1}{c}{Test statistic} & \multicolumn{1}{c}{\textit{p} value} \\
\midrule
\multicolumn{5}{l}{\textit{OLS - w/o covariates - X = 0}} \\
(Intercept) $\left(\alpha_0\right)$& 49.3626 & 0.0590 & 836.40 & $< 0.001$ \\ 
SMART $\left(\alpha_1\right)$ & -0.5380 & 0.0440 & -12.24 & $< 0.001$\\ 
$\text{SMART}^2$ $\left(\alpha_2\right)$ & -0.0205 & 0.0060 & -3.42 & $< 0.001$\\ 
\addlinespace
\multicolumn{5}{l}{\textit{OLS - w/o covariates - X = mean based centered}} \\
(Intercept) $\left(\alpha_0\right)$ & 47.6214 & 0.0428 & 1111.62 & $< 0.001$ \\ 
SMART $\left(\alpha_1\right)$  & -0.6577 & 0.0138 & -47.50 & $< 0.001$ \\ 
$\text{SMART}^2$ $\left(\alpha_2\right)$ & -0.0205 & 0.0060 & -3.42 & $< 0.001$ \\ 
\addlinespace
\multicolumn{5}{l}{\textit{OLS - w covariates - X = 0}} \\
(Intercept) $\left(\alpha_0\right)$ & 46.2988 & 0.0732 & 632.70 & $< 0.001$ \\ 
SMARTc $\left(\alpha_1\right)$ & -0.0219 & 0.0432 & -0.51 & 0.61 \\ 
$\text{SMARTc}^2$ $\left(\alpha_2\right)$ & -0.0579 & 0.0059 & -9.86 & $< 0.001$ \\ 
GENDER $\left(\alpha_3\right)$ & 4.5942 & 0.0553 & 83.08 & $< 0.001$ \\ 
DEPRIVED $\left(\alpha_4\right)$ & -0.4223 & 0.0558 & -7.56 & $< 0.001$ \\ 
MINORITY $\left(\alpha_5\right)$ & 0.3011 & 0.0651 & 4.62 & $< 0.001$ \\ 
\addlinespace
\multicolumn{5}{l}{\textit{OLS - w covariates - X = mean based centered}} \\
(Intercept) $\left(\alpha_0\right)$ & 45.7439 & 0.0535 & 854.60 & $< 0.001$ \\ 
SMARTc $\left(\alpha_1\right)$ & -0.3592 & 0.0139 & -25.79 & $< 0.001$ \\ 
$\text{SMARTc}^2$ $\left(\alpha_2\right)$ & -0.0579 & 0.0059 & -9.86 & $< 0.001$ \\ 
MALE $\left(\alpha_3\right)$ & 4.5942 & 0.0553 & 83.08 & $< 0.001$ \\ 
DEPRIVED $\left(\alpha_4\right)$ & -0.4223 & 0.0558 & -7.56 & $< 0.001$ \\ 
MINORITY $\left(\alpha_5\right)$ & 0.3011 & 0.0651 & 4.62 & $< 0.001$ \\ 
\addlinespace
\multicolumn{5}{l}{\textit{PIM - w/o covariates}} \\
SMART $\left(\beta_1\right)$& -0.05 & 0.0034 & -14.05 & $< 0.001$ \\ 
$\text{SMART}^2$ $\left(\beta_2\right)$ & -0.0005 & 0.0005 & -1.15 & 0.25 \\ 
\addlinespace
\multicolumn{5}{l}{\textit{PIM - w covariates}} \\  
SMART $\left(\beta_1\right)$ & -0.008 & 0.0036 & -2.30 & 0.02 \\ 
$\text{SMART}^2$ $\left(\beta_2\right)$ & -0.004 & 0.0005 & -7.53 & $< 0.001$ \\ 
MALE $\left(\beta_3\right)$ & 0.365 & 0.0044 & 82.21 & $< 0.001$ \\ 
DEPRIVED $\left(\beta_4\right)$ & -0.034 & 0.0044 & -7.64 & $< 0.001$ \\
MINORITY $\left(\beta_5\right)$ & 0.017 & 0.0052 & 3.32 & $< 0.001$ \\ 
\addlinespace
\bottomrule
\end{tabular} 
  \caption{Results of the models with squared predictor using \gls{ols} without mean based centering or with mean based centering. Below are results for fitting the \gls{pim}. Each model is fitted without (w/o) and with (w) covariates. The test statistics correspond to $t$-values with  for \gls{ols} and $z$-statistics for \gls{pim}. Degrees of freedom for $t$-values in the model without covariates = 116627. With covariates = 116624.} \label{table:quad_results} 
\end{table} 

Finally we also fit model (\ref{pim_noc_quad}) without and with covariates. Estimates are given in the lower part of Table \ref{table:quad_results}. We observe a significant negative effect of both the linear ($\beta_1 = -0.008, z =  -2.30, p = 0.02$) and quadratic ($\beta_2 = -0.004, z =  -7.53, p < 0.001$) predictor on the \gls{pi} of mental well-being when covariates are included in the model.

\subsubsection{ANOVA}
So far, the results for both the linear regression models and the \gls{pim}s may seem unsatisfying. Indeed, while the average reported mental well-being initially increases, our estimates for the average mental well-being and the \gls{pi} show decreasing patterns only. It is possible to include an interaction term between $X$ and $X^2$, though we explore a simple alternative in this section. Our suggestion can be applied to both the linear regression model and the \gls{pim}, though we will restrict the analysis to the latter. \\

\begin{figure}[!h]
\begin{center}
\includegraphics[width=0.75\textwidth]{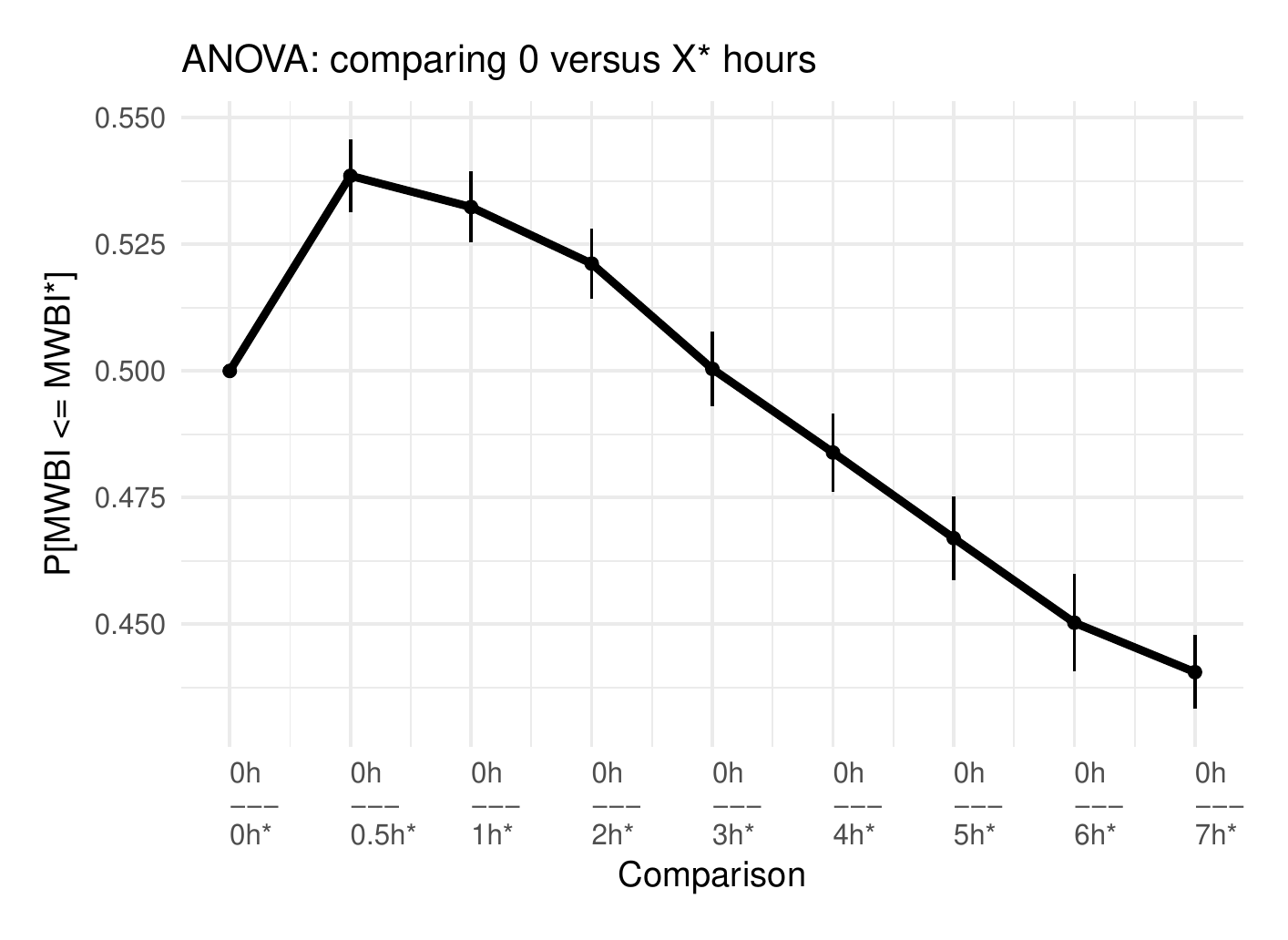}
\caption{Estimated probabilities of a randomly chosen adolescent who does not use a smartphone reporting a lower mental well-being when compared with a same aged randomly chosen adolescent who engages X hours with their smartphone during the week. Effects are controlled for gender, ethnic background and economic background.}\label{anova_plot}
\end{center}
\end{figure}

In previous models, we have used the predictor $X$ as a continuous variable. It is possible however to allow for more flexibility by coding the Likert scale of time spent with a smartphone as a factor with 9 levels. We thus have observations $Y_j$ and a predictor $X_j$ with $j \in \{1, \dots, 9\}$ corresponding to $0, 0.5, 1, 2, 3, 4, 5, 6 \text{ or } 7$ hours of smartphone usage. There are two implications for this. First, we do not assume equidistant intervals any more (i.e. the effect between 0-1 hours is not equal any more as the effect between 1-2 hours). Second, this approach does not allow for generalizations beyond 7 hours of time spent with smartphones. The advantage is that we can compare each level with a baseline, which we set to $X_1 = 0$ hours smartphone usage. \\
To see, consider that 8 binary dummy variables will be created for each $X_j$ with $j > 1$. Each of these binary variables encodes group membership for $0.5, 1, 2, 3, 4, 5, 6 \text{ or } 7$ hours. The corresponding \gls{pim} is then
\begin{align}  \nonumber 
\text{probit} \{\text{P}(MWBI \preceq  MWBI^{*}) \} = &\beta_1(X^*_2 - X_2) + \beta_2(X^*_3 - X_3) + \\ \nonumber
& \beta_3(X^*_4 - X_4) + \beta_4(X^*_5 - X_5) + \\ \nonumber
&\beta_5(X^*_6 - X_6) +\beta_6(X^*_7 - X_7) + \\ 
&\beta_7(X^*_8 - X_8) + \beta_8(X^*_9 - X_9). \label{anova_pim}
\end{align}
Now when we compare a randomly chosen adolescents who uses his/her smartphone for 0.5 hours (denoted as $Y_2$ with $X_2 = 1$) with a randomly chosen adolescent from the baseline group (i.e. 0 hours, $Y_1$), we get
\begin{align}  \nonumber 
\text{probit} \{\text{P}(MWBI_1 \preceq  MWBI_2^{*}) \} = &\beta_1(X^*_2 - X_2) \\  \nonumber 
=& \beta_1
\end{align}
and similar for the other parameters. Since \gls{pim} (\ref{anova_pim}) is of standard form, we can use the \textit{single data partitioning} algorithm. We also extend \gls{pim} (\ref{anova_pim}) with the covariates. Using 4 cores, it took $1.44183$ minutes to fit model \ref{anova_pim} without the covariates and $1.876509$ minutes with covariates. Results of the \gls{pim} fits are given in Table \ref{table:anova_results}. \\
We now observe a probability of $\Phi(\beta_1) = 53.98\%\ [95\% \text{ CI: } 53.131\%; 54.57\%]$ of adolescents reporting a \textit{higher} mental well-being when on average they use their smartphone for half an hour compared to adolescents drawn at random who do not use a smartphone during the week. On the other hand, a randomly chosen adolescent who spends 7 hours with his/her smartphone will have a probability of $1 - \Phi(\beta_8) = 55.96\%\ [95\% \text{ CI: }55.22; 56.68\%]$ of reporting a \textit{lower} mental well-being compared to a randomly chosen adolescent who does not use a smartphone. These percentages are controlled for the effect of gender, economic background and ethnic background. Note that there is no significant effect between 0 and 3 hours usage of a smartphone.\\
A plot with all $\text{P}(MWBI \preceq  MWBI^{*})$ for model (\ref{anova_pim}) with the covariates is shown in Figure \ref{anova_plot}. \\

\begin{table}[!ht] \centering 
\captionsetup{width=0.9\textwidth}
\begin{tabular}{lcccc} 
\toprule
\multicolumn{1}{c}{Parameter} & \multicolumn{1}{c}{Estimate} & \multicolumn{1}{c}{Standard error} & \multicolumn{1}{c}{Z-values} & \multicolumn{1}{c}{\textit{p} value} \\
\midrule
\multicolumn{5}{l}{\textit{ANOVA PIM - w/o covariates}} \\
$0 < 0.5\ (\beta_1) $ &  0.07 & 0.0091 & 8.09 & $< 0.001$ \\
$0 < 1\ (\beta_2) $ & 0.038 & 0.0088 & 4.30 & < 0.001 \\ 
$0 < 2\ (\beta_3) $ & -0.017 & 0.0088 & -1.95 & 0.05 \\
$0 < 3\ (\beta_4) $ & -0.093 & 0.0092 & -10.10 & $ < 0.001$ \\ 
$0 < 4\ (\beta_5) $ & -0.155 & 0.0097 & -15.93 & $ < 0.001$ \\
$0 < 5\ (\beta_6) $ & -0.211 & 0.01047 & -20.17 & $ < 0.001$ \\
$0 < 6\ (\beta_7) $ & -0.260 & 0.0123 & -21.12 & $ < 0.001$ \\
$0 < 7\ (\beta_8) $ & -0.284 & 0.0092 & -30.92 & $ < 0.001$ \\
\addlinespace
\multicolumn{5}{l}{\textit{ANOVA PIM - w covariates}} \\
$0 < 0.5\ (\beta_1) $ & 0.100 & 0.0092 & 10.47 & $ < 0.001$ \\
$0 < 1\ (\beta_2) $ &  0.081 & 0.0090 & 9.01 & $ < 0.001$ \\ 
$0 < 2\ (\beta_3) $ & 0.035 & 0.0090 & 5.93 & $ < 0.001$ \\ 
$0 < 3\ (\beta_4) $ & 0.001 & 0.0094 & 0.10 & 0.92 \\ 
$0 < 4\ (\beta_5) $ & -0.040 & 0.0099 & -4.06 & $ < 0.001$ \\ 
$0 < 5\ (\beta_6) $ & -0.083 & 0.0107 & -7.77 & $ < 0.001$ \\ 
$0 < 6\ (\beta_7) $ & -0.125 & 0.0125 & -10.01 & $ < 0.001$ \\ 
$0 < 7\ (\beta_8) $ & -0.150& 0.0094 & -15.86 & $ < 0.001$ \\ 
GENDER $\left(\beta_9\right)$ & 0.3668 & 0.0045 & 82.26 & $ < 0.001$ \\ 
DEPRIVED $\left(\beta_{10}\right)$ & 0.0159 & 0.0052 & 3.04 & 0.002 \\ 
MINORITY $\left(\beta_{11}\right)$ & -0.0317 & 0.0045 & -7.11 & $ < 0.001$ \\ 
\addlinespace
\bottomrule
\end{tabular} 
  \caption{Results of fitting the \gls{pim} \ref{anova_pim} and extending the same model with covariates.} \label{table:anova_results} 
\end{table}

	\chapter{Discussion} \label{Chapter:Disc}

\section{Simulation study}
In the first part of our simulation study, we used a \textit{single data partitioning} algorithm to obtain estimates for a \gls{pim} when the sample size is large. This algorithm consists of subdividing the entire data set into non-overlapping partitions. On each of those, the \gls{pim} is fitted and estimates are pooled. We observed favourable results as the final \gls{pim} estimator is unbiased, approximately normally distributed and has empirical coverages of the confidence intervals around nominal level $1 - \alpha$, for both solutions of calculating the variance of the estimator. \\

In the second part of our simulation study, we used a \textit{uniform subsampling} algorithm for the same purpose. For $B$ iterations, we sampled $K \ll n$ original observations without replacement and obtained \gls{pim} estimates on these subsets. To calculate the variance of our estimator, we used a scaling approach as well as adjusting the sandwich variance estimator. We showed how the estimates are converging to the true value of $\beta$ when the amount of observations and resampling iterations increase. Also, we are able to fit a \gls{pim} in a matter of seconds when $K$ and $B$ = 230. However, we were unable to achieve desirable empirical coverages of the 95\% \gls{ci}s. We discuss two findings regarding these coverages. \\
First, for the scaled variance, we rely for a \gls{pim} with a one-dimensional predictor on the following property:
\begin{align} \label{convergenge}
\sqrt{\frac{K}{n}}(\hat{\beta}^{*} - \beta) \xrightarrow{D} N(0, \sigma^{2})
\end{align}
as $n,K \rightarrow +\infty$. Note that we use the sampling distribution of $\beta^*$ which is based on the subsample $K \ll n$. Hence we do not use the number of pseudo-observations, but the amount of original data points. We assume this distribution is the same as the sampling distribution of the \gls{pim} estimator on the full data set since we sample without replacement. Hence asymptotic theory suggest we should get a convergence given by equation (\ref{convergenge}). Our results for the \textit{scaled \gls{ci}} do seem to converge to 0.95 as $K$ increases. However the rate of convergence, which depends on $\sqrt{K/n}$ is too slow for our purpose. We checked in a quick simulation the empirical coverages of 95\% \gls{ci}s for the parameters obtained through \gls{ols} in a \textit{uniform subsampling} algorithm with the same calculations for the \textit{scaled \gls{ci}}. We observed a similar slow rate of convergence. Moreover, by letting $K$ go to 2500 (instead of 1000), we see the coverage slowly reaching 0.95. Result is given in the appendix (Figure \ref{appendix:convergence}). \\
Second, we observe how the empirical coverages using the \textit{adjusted sandwich variance estimators} approach 0.95, but then decrease again. We can explain these results by looking at the sampling algorithm itself. Note that we sample observations with replacement between iterations, though without replacement within iterations. Hence as the amount of subsampling iterations increases, it becomes more likely that estimates between iterations are correlated. However, we assumed that the covariance between those estimates is equal to 0. This also explains the observed coverage of 0.95 when 10 iterations together with 1000 observations are used. \\

We list two disadvantages of the \textit{single data partitioning} algorithm compared to uniform subsampling. First it takes (slightly) longer to compute, depending on the amount of cores available. One will also need more cores if the sample size increases even further. Although this is true for every statistical model, \textit{uniform subsampling} permits a user to have a quick look at the data without defining a function that can be computed in parallel over different cores. \\
Second, one should theoretically check whether results are depending on the data partitioning as this algorithm does not compare all observations with each other. \\
In our results however, we concluded that the statistical performance of this algorithm outweighs its disadvantages listed above. \\

\section{Optimal resampling}
One potential area for improvement in the current \textit{uniform subsampling} algorithm are the subsampling probabilities $\pi = \{\pi_i\}_{i=1}^n$ assigned to each observation. In research areas involving big data, it is becomes hard to fit simple models such as linear or logistic regression models \citep{Wang2016} on the full data set. One can deploy a subsampling algorithm, as is done here. However, computational efficiency can be increased by sampling with replacement using non-uniform subsampling probabilities. The key is to sample those observations which are highly influential. For instance, \cite{Ma2015} first obtain leverage scores through singular value decomposition and use these as subsampling probabilities. Based on the work of \cite{Drineas2006, Drineas2011}, they provide an unbiased estimator using weighted least squares regression on the subset. Likewise, \cite{Wang2017} derive optimal subsampling probabilities which minimize the variance of the estimator for logistic regression. This is based on the A-optimality criteria from optimal designs of experiments (i.e. using the trace of the inverse of the information matrix). \\
In preparation of this thesis, we encountered two challenges when implementing a similar subsampling algorithm for \gls{pim}s. First one should be able to define influential observations. As the \gls{pim} estimator is related to several other statistical models \citep{Thas2012, DeNeve2015}, it could be possible to construct subsampling probabilities using a related model. If for instance the normal linear regression holds, such as our data generating models in the Monte Carlo simulations, one could use normalized leverage scores based on \gls{ols}. Second, one needs to know how the subsampling probabilities/weights given to each observation transform to pseudo-observations. This is true as estimating $\beta$ without controlling for the non-uniform sampling process might result in a biased estimate. In earlier stages, we tried to average $W$ and $W^*$ which are the weights given to observation $Y$ and $Y^{*}$ respectively. We then used a weighted score function to estimate $\boldsymbol\beta$:
\begin{align} \label{WEsEq_1}
U^*_n(\boldsymbol{\beta}) = \sum_{(i,j) \in I_n} \mathbf{A}(\mathbf{Z}_{ij};\boldsymbol{\beta}) \times W_{ij} \times [I_{ij} - g^{-1}(\mathbf{Z}_{ij}\boldsymbol{\beta})] = 0 \ \ ,
\end{align}
where $W_{ij}$ is the average leverage score of each pseudo-observation $(i,j) \in I_{n}$. Results however revealed a biased estimator. Two figures containing $\beta$ \gls{pim} estimates and the 95\% \gls{ci}s using the \textit{scaled standard error} are included in the appendix.


\section{Application}
We used a \gls{pim} to provide a natural quantification of the effect of smartphone usage on mental well-being as measured by the Warwick-Edinburgh Mental Well-Being Scale \citep{Tennant2007}. To fit a \gls{pim} on a data set of $116,630$ adolescents from the UK, we used the \textit{single data partitioning} algorithm with 95\% \gls{ci}s based on the \textit{adjusted sandwich estimator}. As the average mental well-being versus time spent during the week is described by a curvilinear pattern, we found the most satisfying results using an ANOVA formulation of the \gls{pim}. We estimated probabilities of a randomly chosen adolescent who does not use a smartphone during the week reporting a lower mental well-being compared to peers. At first this probability increases to 53.98\% when compared to peers who use their smartphone on average for half an hour during the week. This then drops to 44.04\% when compared to peers who use their smartphone for 7 hours. The effect is controlled for variables such as gender, ethnic background and living in wealthy areas. It took between 1 and 2 minutes to fit this model using a single machine with 4 cores. \\

Some additional comments are provided here. First while we have validated the \textit{single data partitioning} algorithm in a simulation study, we used $n = 250.000$ which leads to 1000 partitions of size 2500. It remains unclear how this algorithm performs when the amount and size of the partitions shrink. In our application, we have 117 partitions of size 1000. Even though we expect similar performances, one could easily set up a simulation study bridging the gap between $n = 2500$ and $n = 250.000$ or beyond. \\
Second, we have not provided formal tests for the goodness-of-fit \citep{DeNeve2013a} of our \gls{pim}s (nor appropriate graphical tools). As stated earlier, assessing these requires again computational improvements. Finally, there is still a possibility to evaluate other methods such as bootstrapping \textit{m out of n} observations in future research.

	\chapter{Conclusion}

In this dissertation we have demonstrated how to fit Probabilistic Index Models on data sets where standard estimation procedure is computationally not possible. We have evaluated two algorithms to do so. One consists of creating unique \textit{partitions} of the original data set, fit the model onto each partition and combine estimates. The second algorithm consists of iterating a subsampling scheme (without replacement) in which all observations have equal subsampling probability. We used two approaches of calculating the variance of the resulting estimator. One based on \textit{scaling} the standard error and one based on adjusted the sandwich variance estimator. \\
Our results favour the usage of the \textit{single data partitioning} algorithm. We have then used this algorithm to estimate the Probabilistic Index on an existing data set ($n = 116,630$). We were able to fit a Probabilistic Index Model written as an ANOVA fairly rapidly and provide meaningful effect sizes (see Figure \ref{anova_plot}). This is given as alternative to the ordinary least squares procedure in normal linear regression models.

\section*{Acknowledgements}
The computational resources (Stevin Supercomputer Infrastructure) and services used in this work were provided by the VSC (Flemish Supercomputer Center), funded by Ghent University, the Hercules Foundation, and the Flemish Government department EWI.
	

\phantomsection
\addcontentsline{toc}{chapter}{References}
\renewcommand{\bibname}{References}
\bibliographystyle{humannat}
\bibliography{DB_THESIS}
	\part*{Appendices}
\appendix

\chapter{Normal QQ plots}
\clearpage
\section{Data generating model 1}
\begin{figure}[!h]
\begin{center}
\includegraphics[width=0.90\textwidth]{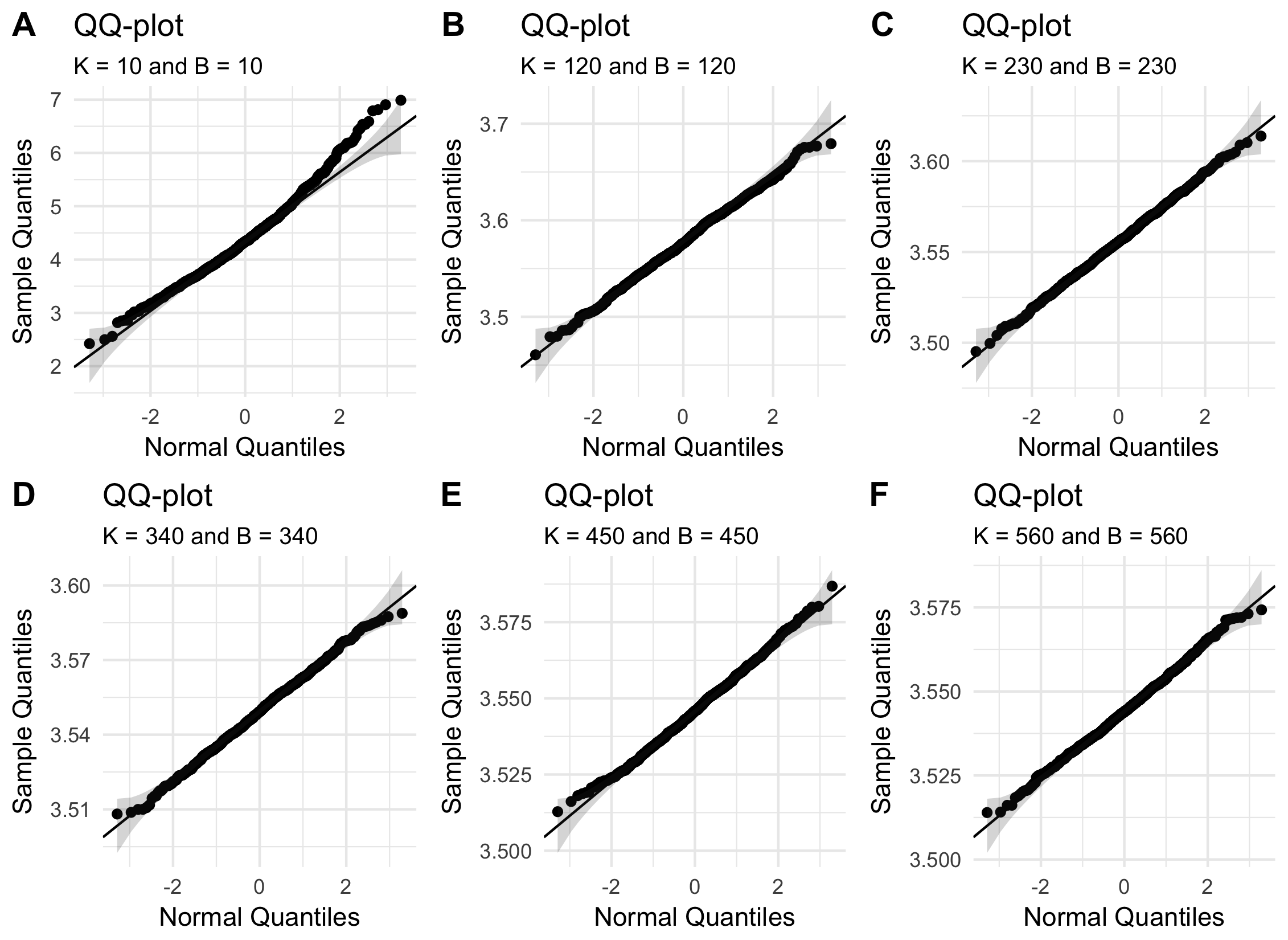}
\includegraphics[width=0.90\textwidth]{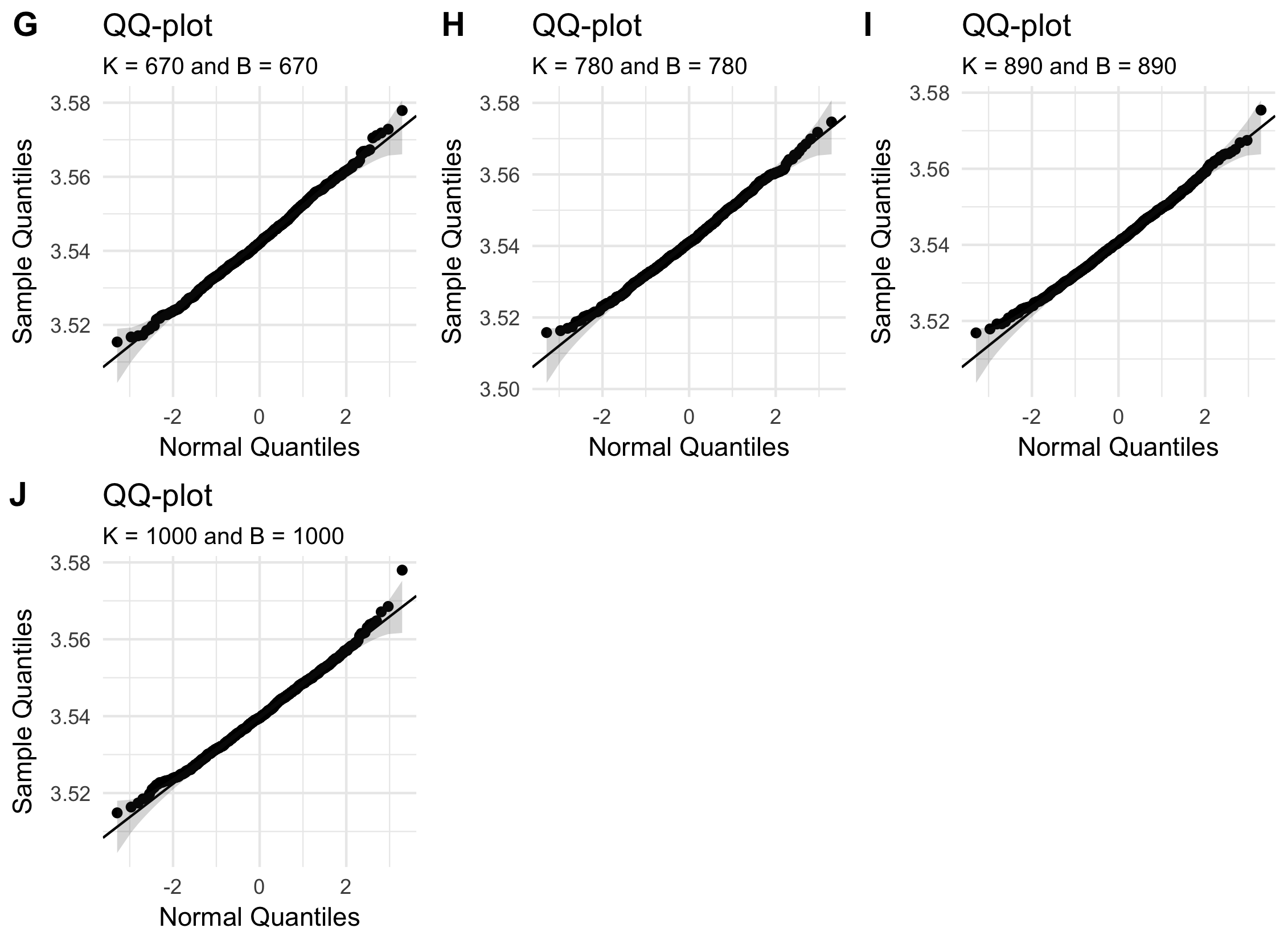}
\caption{Normal QQ-plots of $\hat\beta$ for all $K$ and $B$  in \textit{uniform subsampling algorithm}. Data is generated under model 1 (Section \ref{data_model1}). }\label{appendix:QQ_mod1}
\end{center}
\end{figure} 

\clearpage

\section{Data generating model 2}

\begin{figure}[!h]
\begin{center}
\includegraphics[width=0.90\textwidth]{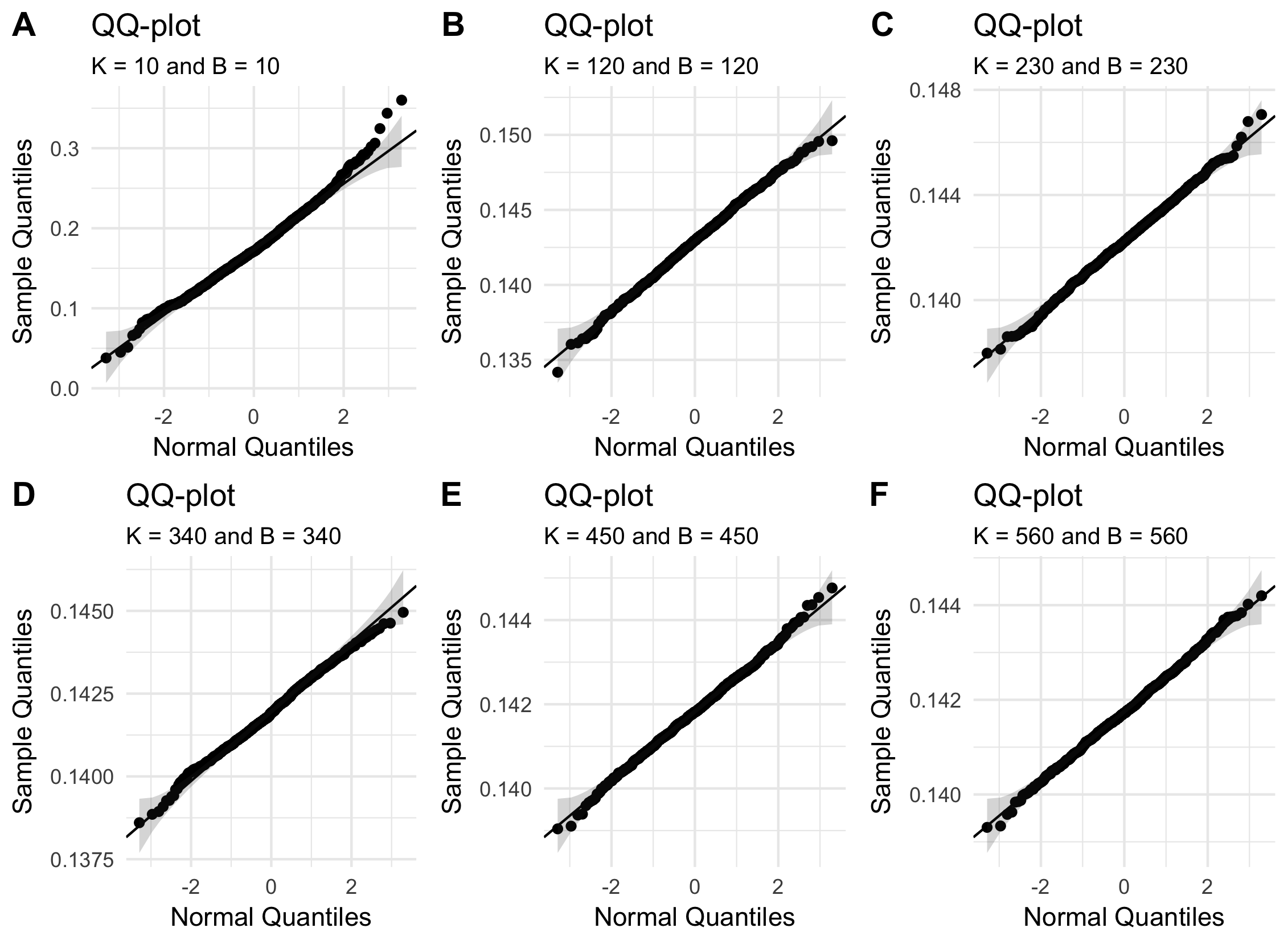}
\includegraphics[width=0.90\textwidth]{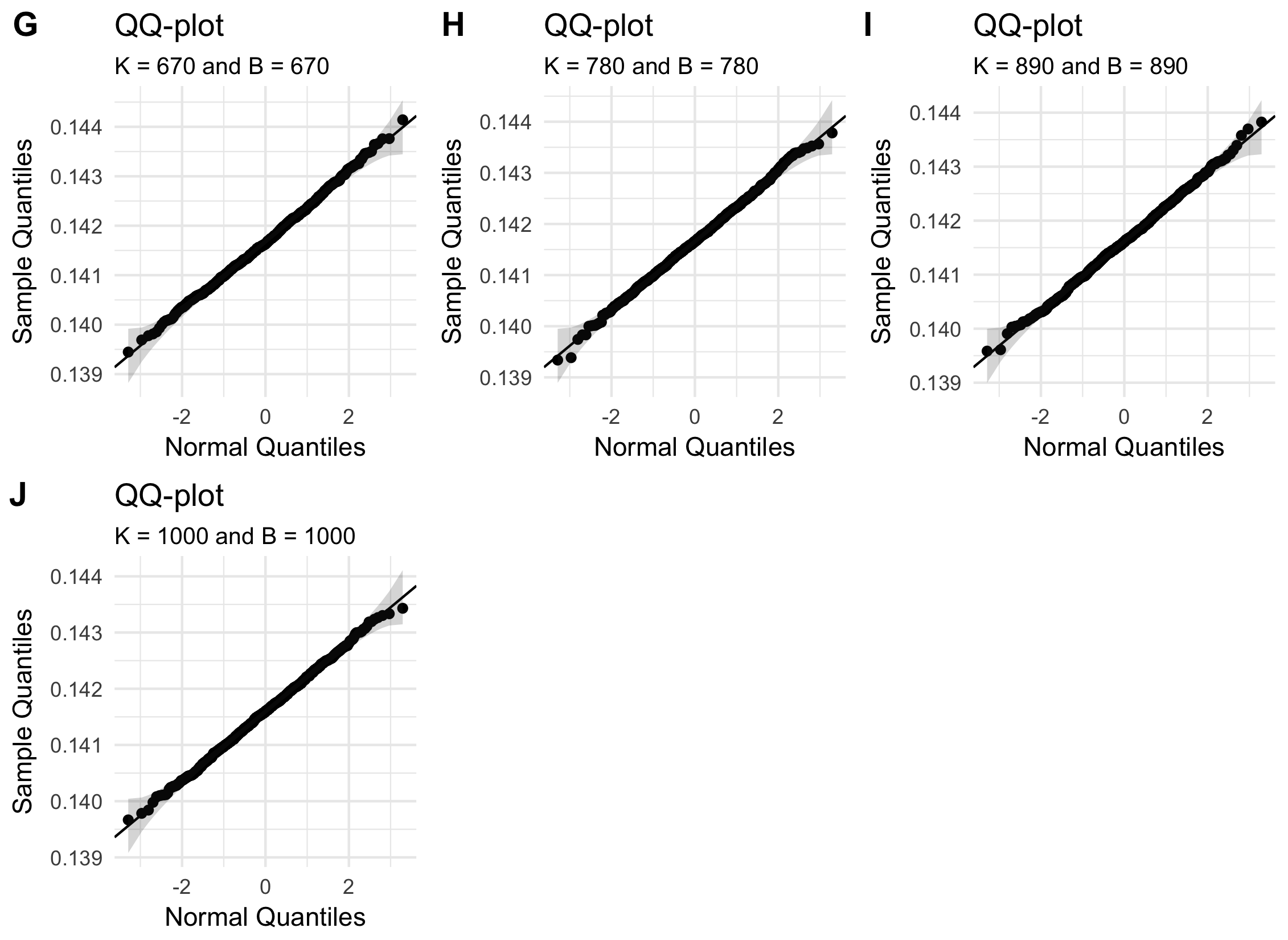}
\caption{Normal QQ-plots of $\hat\beta$ for all $K$ and $B$  in \textit{uniform subsampling algorithm}. Data is generated under model 2 (Section \ref{data_model2}).}\label{appendix:QQ_mod2}
\end{center}
\end{figure} 

\clearpage
\section{Data generating model 3}

\begin{figure}[!h]
\begin{center}
\includegraphics[width=0.90\textwidth]{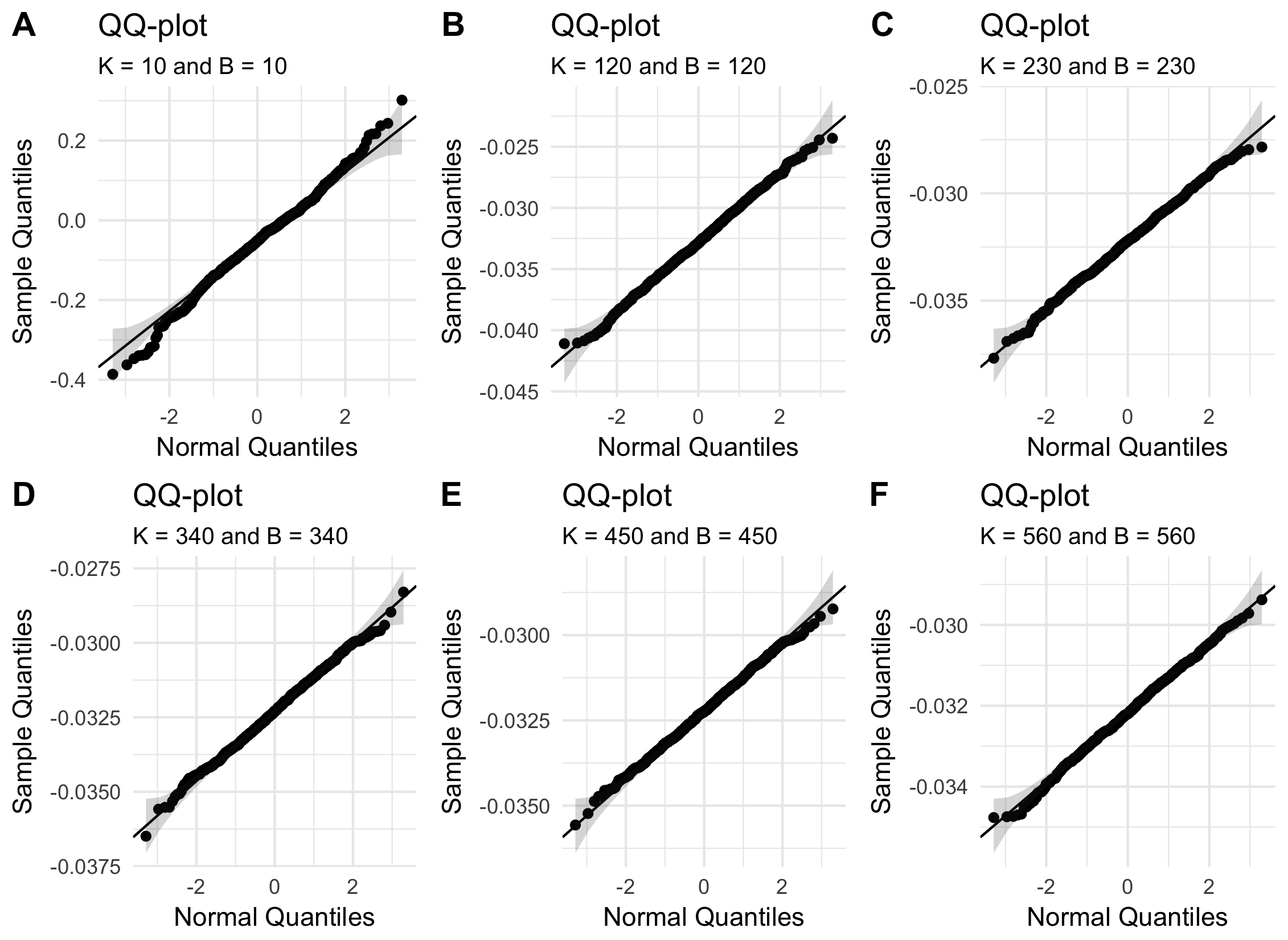}
\includegraphics[width=0.90\textwidth]{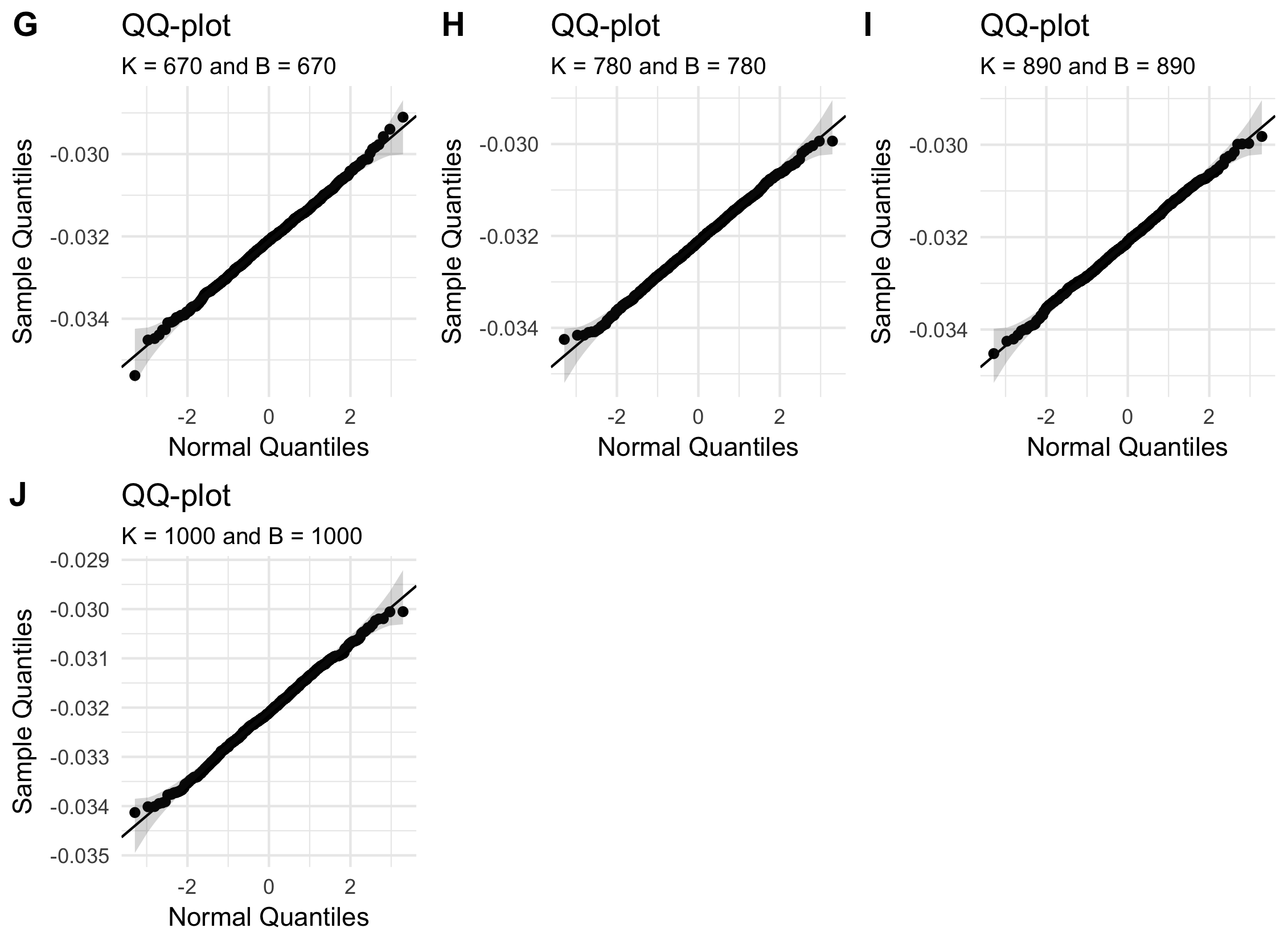}
\caption{Normal QQ-plots of $\hat\beta$ for all $K$ and $B$  in \textit{uniform subsampling algorithm}. Data is generated under model 3 (Section \ref{data_model3}).}\label{appendix:QQ_mod3}
\end{center}
\end{figure} 

\clearpage

\chapter{Convergenge}
\begin{figure}[!h]
\begin{center}
\includegraphics[width=0.80\textwidth]{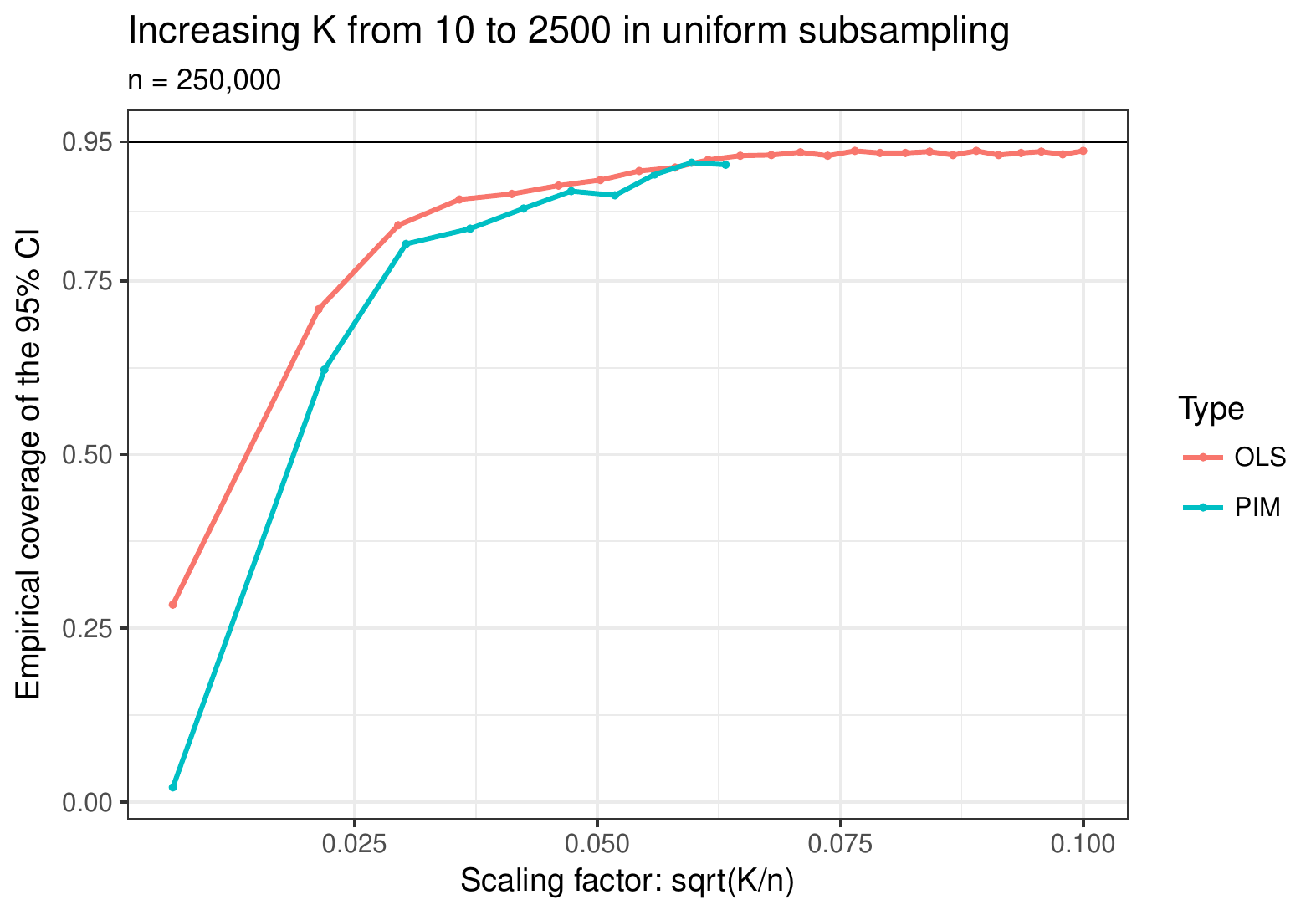}
\caption{Empirical coverage of the 95\% \gls{ci} for parameters in a univariate linear regression model ($\alpha$) using \gls{ols} and a \gls{pim} ($\beta$) in a uniform subsampling context. On x-axis: scaling factor for the \textit{scaled \gls{ci}} of equation (\ref{scaled_ci_uniform_sampling}). For 1000 iterations, $K$ observations are sampled from $n = 250,000$. We let $K$ vary from 10 to 2500 for \gls{ols}. We plotted the results of the \gls{pim} estimator which goes to $K = 1000$ as a reference. Based on 1000 Monte Carlo simulation runs.} \label{appendix:convergence}
\end{center}
\end{figure}

\chapter{Non uniform subsampling}

\begin{figure}[!h]
\begin{center}
\includegraphics[width=0.75\textwidth]{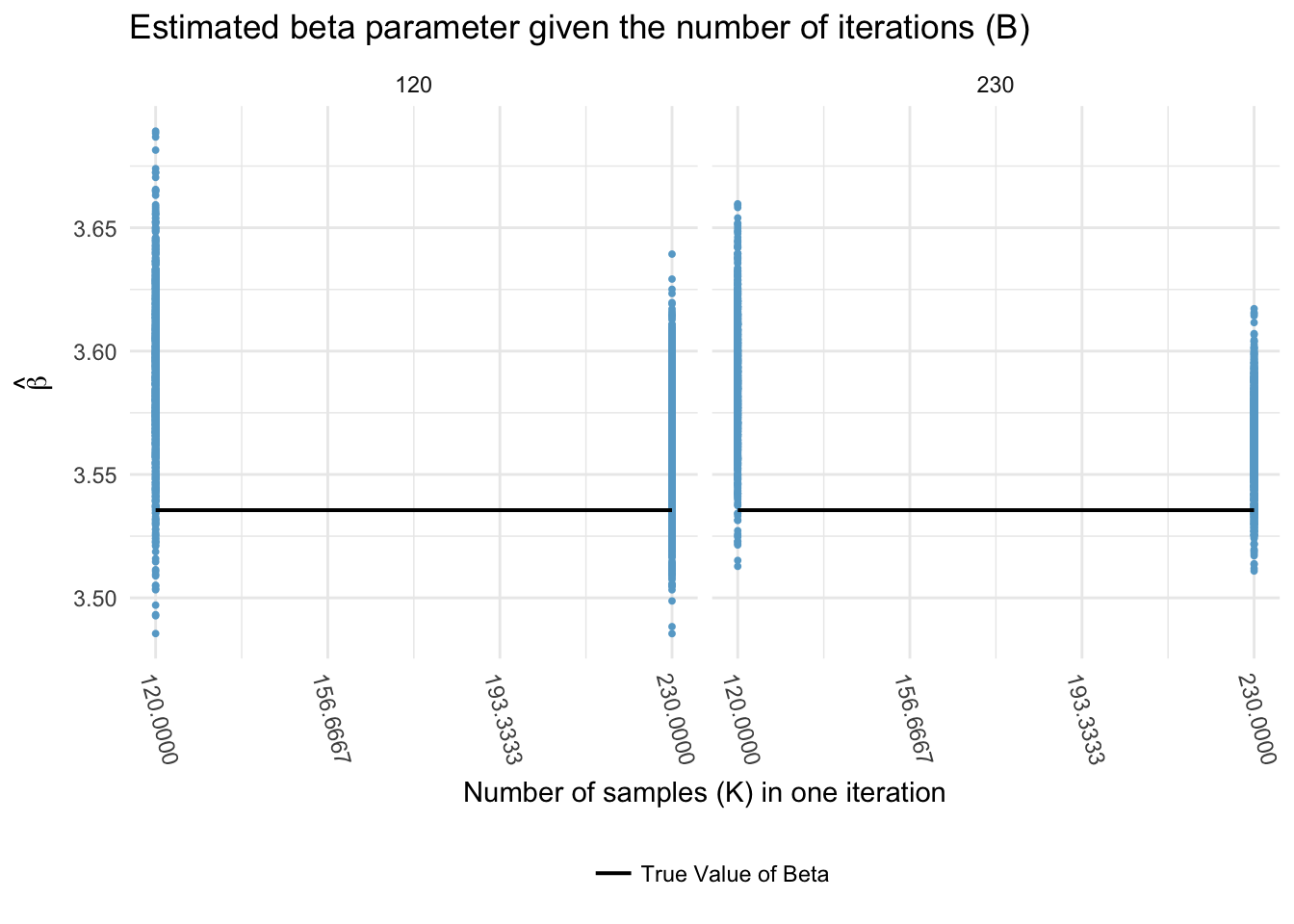}
\caption{Simulation results for 1000 Monte Carlo runs using a non-uniform sampling algorithm. Weights for original observations are determined using normalized leverage scores and averaged to get weighted pseudo-observations. Estimation through weighted score function. Data is generated under model 1 (Section \ref{data_model1}). Panel contains $\beta$ \gls{pim} estimates with respect to the true parameter.}\label{appendix:nonUniformBeta}
\end{center}
\end{figure}

\begin{figure}[!h]
\begin{center}
\includegraphics[scale = 0.3]{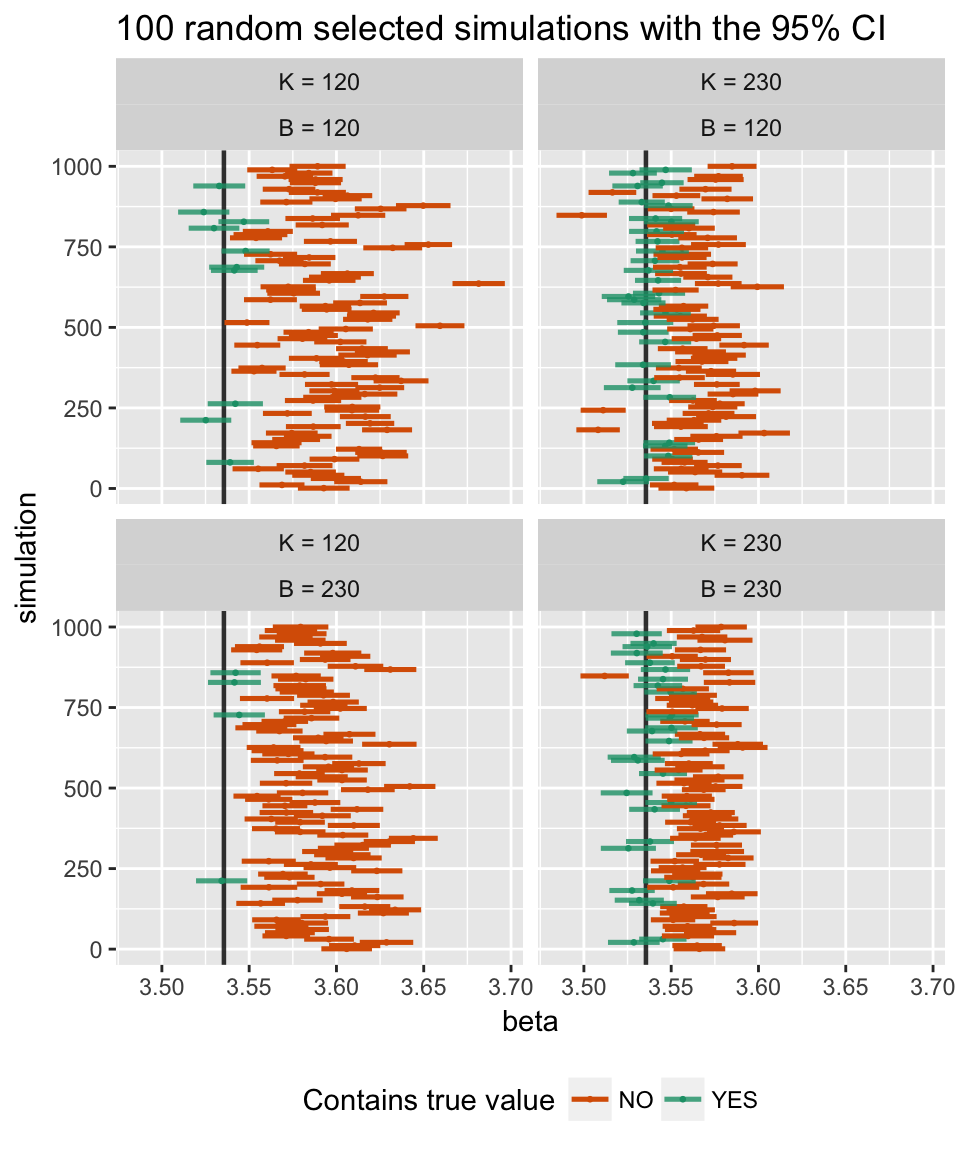}
\caption{Simulation results for 1000 Monte Carlo runs using a non-uniform sampling algorithm. Weights for original observations are determined using normalized leverage scores and averaged to get weighted pseudo-observations. Estimation through weighted score function. Data is generated under model 1 (Section \ref{data_model1}). Panel shows 100 at random selected 95\% \gls{ci} for $\beta$ using the \textit{scaled standard error}.}\label{appendix:ci}
\end{center}
\end{figure}  

\end{document}